**Genomic Informational Field Theory (GIFT) to characterize genotypes involved in large phenotypic fluctuations.**

Cyril Rauch[1], Panagiota Kyratzi[2] and Andras Paldi[2]

[1] School of Veterinary Medicine and Science, University of Nottingham, College Road, Sutton Bonington, LE12 5RD, United Kingdom

[2] École Pratique des Hautes Études, PSL Research University, St-Antoine Research Center, Inserm U938, 34 rue Crozatier, 75012, Paris, France

**Email for correspondence**: cyril.rauch@nottingham.ac.uk.

**Abstract**

Based on the normal distribution and its properties, i.e., average and variance, Fisher works have provided a conceptual framework to identify genotype-phenotype associations. While Fisher intuition has proved fruitful over the past century, the current demands for higher mapping precisions have led to the formulation of a new genotype-phenotype association method a.k.a. GIFT (Genomic Informational Field Theory). Not only is the method more powerful in extracting information from genotype and phenotype datasets, GIFT can also deal with any phenotype distribution density function. Here we apply GIFT to a hypothetical Cauchy-distributed phenotype. As opposed to the normal distribution that restricts fluctuations to a finite variance defined by the bulk of the distribution, Cauchy distribution embraces large phenotypic fluctuations and as a result, averages and variances from Cauchy-distributed phenotypes cannot be defined mathematically. While classic genotype-phenotype association methods (GWAS) are unable to function without proper average and variance, it is demonstrated here that GIFT can associate genotype to phenotype in this case. As phenotypic plasticity, i.e., phenotypic fluctuation, is central to surviving sudden environmental changes, by applying GIFT the unique characteristic of the genotype permitting evolution of biallelic organisms to take place is determined in this case.

**Introduction**

A central question in biology is to understand how genetic variation contribute to phenotypic variation in a population. The seminal works by R.A. Fisher [1], [2] published early in the 20[th] century, aimed at settling a fierce debate between Mendelians and Biometricians linked to the apparent disjunction between the so-called Mendelian (monogenic) factors leading to discrete phenotypic changes and the fact that most natural phenotypes are continuous. Fisher used the field of frequentist probability (i.e., distribution density functions in the continuum limit) and hypothesized that if many genes affect traits, the random sampling of alleles at each gene would produce a continuous and normally distributed phenotype out of which the phenotypic average and variance would have fundamental meaning for genetic.

Since its inception, Fisher's method, a.k.a. method of averages, has been applied to a large range of datasets. The NHGRI-EBI Genome-Wide Association Studies (GWAS) Catalogue listed 316,782 associations identified in 5149 publications [3]. While the results obtained are indeed encouraging, GWASs suffer strong limitations linked to the notion of statistical power, namely the fact that the precision in the inferences relies on the sample size. While this is not so much of a problem when the gene effect/effect size is large, it is significant when complex traits involving numerous variants with small effect size are studied. A good example of such limitation for classic GWAS is the phenotype height in humans. Studied for well over a century, the phenotype height in humans is a model for investigating the genetic basis of complex traits [1], [4] and whose heritability is well known to be around 80% [1], [5], [6]. This complex trait has remained controversial for a long time [7] as GWASs performed on this phenotype were only able to recover at most 60% heritability [8], [9]. For many years this missing heritability was thought to be associated with the restricted sample size used or the involvement of an ill-defined environment [10], [11]. As the notion of heredity is linked to genetic variances, the conclusions suggested also that loci are missing accounting for the missing heritability measured. Using a staggeringly large sample size of 5.4 million individuals, a recent international GWAS claims to have finally produced a saturated map of common genetic variants associated with human height [12]. While producing a saturated map of common genetic variants associated with human height is clearly an achievement in the field of genetic, this study underlines also that small gene effects are remarkably challenging to detect with current statistical models, namely that the time and cost required to map complex traits are exorbitant.

A point to recall here before exploring further the issues surrounding complex traits is that Fisher's method used by GWASs is an empirical method, namely it is a statistical method that is not fundamentally linked to specific scientific principles. A different way to say this is that statistics can be applied to many different and unrelated disciplines because statistics are independent from the said disciplines. Consequently, postulating new empirical methods that are more informative than the previous ones can only be a positive exercise.

Classic GWASs based on Fisher theory are limited in the information they can provide because they use the normal distribution emerging from a theory that considers that information is limited or missing. Indeed, the normal distribution, a.k.a. Law of Errors, as used by Fisher was devised by physicists to understand and model the impact of experimental or observational errors [13]. As physicists deal with the Intemporal Laws of Nature, i.e., determinism, errors in predicting the outcome of an experiment must be linked to experimental imprecisions. As a result, errors occur in physics because the full information on the system studied is incomplete or missing. It follows, as a consequence, that trying to extract precise information from a biological system using the normal distribution, i.e., using a method that considers that information is missing from the start, is deemed to be a challenging exercise. As it turns out, the normal distribution has imposed itself as a universal feature mostly for sociological, economical and political reasons and not scientific ones [14]. However, this presumed universal feature is not always valid mathematically since the normal distribution only emerges when specific assumptions are used [14]. The latter statement is obvious, anyone who has ever generated a bar-chart using real phenotype values will know that a normal distribution hardly ever emerges.

Discussion around the epistemological issues and scientific interpretation of results linked to the use of frequentist probabilities and the normal distribution outside the field of Physics, i.e., in Biology, has already been examined elsewhere [14].

Therefore, as there is no reason to assume that information is missing to map genotype to phenotype, there is no valid reasons to use distribution density functions such as the normal distribution. If density functions are used it is because classic GWA methods assume that the only valuable genotype-phenotype information available is stored in the averages, a.k.a. method of averages, and considers as a result that noise in the data (variance) is required to explain why data is not identical numerically to the calculated average. However, this stance is not always valid as there is a difference between the notion of information given by Shannon's theory of information based on (information) entropy, and the one of average; namely that averages are not always meaningful and informative. It is in this context that we have formulated a new GWA method a.k.a. GIFT that is entirely based on the informational content of datasets, as opposed to concentrating exclusively on their averages. As this new method does not consider the variance in the data as a nuisance, it can map genotype to phenotype very precisely with small sample sizes [15], [16].

While GIFT has been demonstrated to be more efficient than classic GWA/Fisher's method at extracting information from datasets [15], [16], the question we would like to address now is: Can GIFT provide entirely new information on genotype-phenotype mapping that classic GWA/Fisher method would not be able to extract, not because of a population/sample size being not big enough, but because the conceptual frame invented by Fisher is too restricted?

At the conceptual level Fisher theory is fundamentally dependent on the existence and the pertinence of averages to describe genotype-phenotype associations. Since the normal distribution obtained from the central limit theorem is derived using the arithmetic means of data as a pertinent definition of the average [13], [17], using the normal distribution as a template for density functions, as Fisher did, warrants the existence of averages. However, and as said above, the likelihood that upon generating a bar-chart using real phenotype values the normal distribution emerges as a perfect fit is very unlikely. This, in turn, cast doubt concerning the pertinence of using averages to depict or extract information from genotype-phenotype association in every cases.

This issue concerning the choice of using the normal distribution and related average and variance to extract meaningful information from dataset in every case is not new and has been the subject of intense debates. A particularly important debate, in fact the first one, took place during the year 1853 between two brilliant mathematicians, Augustin Louis Cauchy (1789-1857) and Irenee Jules Bienayme (1796-1878) [18]. To make a long story short Augustin Louis Cauchy demonstrated that the Law of Errors from which an interpolation of observational data can be inferred, does not have to be constrained by the principles leading to the normal distribution and he provided another distribution density function, a.k.a. Cauchy distribution, that is devoid of average and variance as a counter example. In short, Augustin Louis Cauchy demonstrated that probability density functions do not need to have well defined average or variance (or any other moments) as constitutive parameters. Today Cauchy distribution belong to a family of very important distributions known as Levy distributions initially defined by Paul Pierre Levy (1897-1971) a.k.a. 'fat tails' distributions. For any further information, a discussion regarding the differences between the normal and Cauchy distributions is relegated in Appendix A. In this context it is clear that if a phenotype were Cauchy distributed, without well-defined averages, classic GWAS would not be able to 'deal' with this type of phenotype to associate genotype.

While this may sound counterintuitive for anyone assuming that the normal distribution is universal and always valid, it is worth recalling that the notion of average in biology remains debated [19], as there does not seem to be any clear way to define the different cellular identities in a population [20], [21]. Furthermore, large phenotypic fluctuations in population as inferred by Cauchy distribution is central to the notion of Evolutionary rescue [22]–[24]. Evolutionary rescue is concerned with a population living in an environment that changes suddenly and that can only survive if some individuals in the population carry a trait that turns out to be successful in the new environment [25]. Contrary to the normal distribution that is too stable to permit frequent large fluctuations, Cauchy-distributed phenotypes would promote 'survival' traits resulting from large fluctuations. Unsurprisingly, large transcriptomic fluctuations as underscored by Cauchy distribution participate to the evolutionary dynamics of cancer cells [26] and appears in living organisms in transient environments [27].

It is in a context where large phenotypic fluctuations in a population are possible that we want to use GIFT and show that the association between the genotype and the phenotype leads to a unique selection of genetic microstates.

While the current manuscript is written in theoretical terms, a lot of efforts have been spent to make the manuscript as understandable as possible. Some reader may find the manuscript rather 'dry'. Although this 'feeling' is understandable, the paradigmatic change that this work suggests is too important to ignore and need to be dealt with accordingly.

Accordingly, the manuscript is organised as follows: In the first part a recapitulation, or review, of GIFT as a method will be laid out to provide the necessary ideas and underlying intuition to become familiar with the method and to perform genotype-phenotype mapping analyses. This first part should be accessible to anyone. In the second part the mathematical formalisation of GIFT will concentrate on the conservation of genetic microstates. This second part is strongly connected to physics field theory and statistical physics and a reader with some notions of physics should understand its development. Finally, the conservation of genetic microstates is an important relation that will be used in a third part to determine the properties of the genotype involved in a population whose phenotype is Cauchy distributed. This last part requires some knowledge of complex analysis and in particular of Cauchy integral theorem (it is the same Cauchy who devised the distribution devoid of average and variance mentioned above). However, the 'hard work' has been relegated to appendices for anyone interested in its development.

1. **Intuitive introduction to the method GIFT**
    1.1. **Impact of precision of phenotypic measurements: Deconstructing GWAS**

GIFT is a method designed to deconvolve the notion of precision of phenotype measurements with the one of sample size. Said differently it is a method that aims to provide genotype-phenotype associations without using the notion of categories/data grouping. To develop the theory underscoring GIFT, it is essential to visualise what means increasing precision in phenotypic measurements with current methods. Let us consider a particular single nucleotide polymorphism (SNP) in the genome that one shall assume is associated with the phenotype considered. One shall call SNP1 this single nucleotide polymorphism. With current GWA methods based on Fisher's theory, the phenotype distribution density function is decomposed over the distribution density function of different genetic microstates to determine genetic inferences regarding genotype-phenotype associations (Fig.1A-top and Fig.1B-top). However, when precision in the phenotypic measurements is increased without changing the sample size (fixed at 1000 individuals in this simulation Fig.1), the distribution density functions are transformed into barcodes (Fig.1A-bottom and Fig.1B-bottom). What Fig.1A and Fig.1B tell us is that deconvolving sample size and precision is equivalent to providing an understanding of the overall configuration of the different barcodes/microstates.

By considering the barcodes, the first thing to note is then that the exact positioning of barcodes in the space of phenotypic values, i.e., the exact positioning of barcodes on the x-axis in Fig.1A-bottom, is a function of the sampling done, namely that different individuals with different phenotypic values within the initial categories would have provided different positionings in the space of phenotypic values. Consequently, what matters here is not the absolute positioning of barcodes but their relative positioning, namely their ordering. Recall that such relative positioning results from having measured the phenotype finely enough, i.e., with very high precision, such that any two individuals from the sampled population will always have different phenotypic values.

In this context, one can then represent the configuration of barcodes as a string of microstates. The position in the string of barcodes is reminiscent of our ability to rank the phenotype values measured as a function of the uniqueness of their magnitude. As the barcodes in Fig.1B are linked to the fact that SNP1 is associated with the phenotype, the key information is therefore contained in the configuration of the string of microstates obtained from Fig.1B-bottom and given by,

**SNP1: [+1, +1, +1, +1, 0, +1, +1, ..., +1, 0, 0, 0, -1, 0, -1, ..., -1, 0, -1, -1, -1, -1, -1]**     (1)

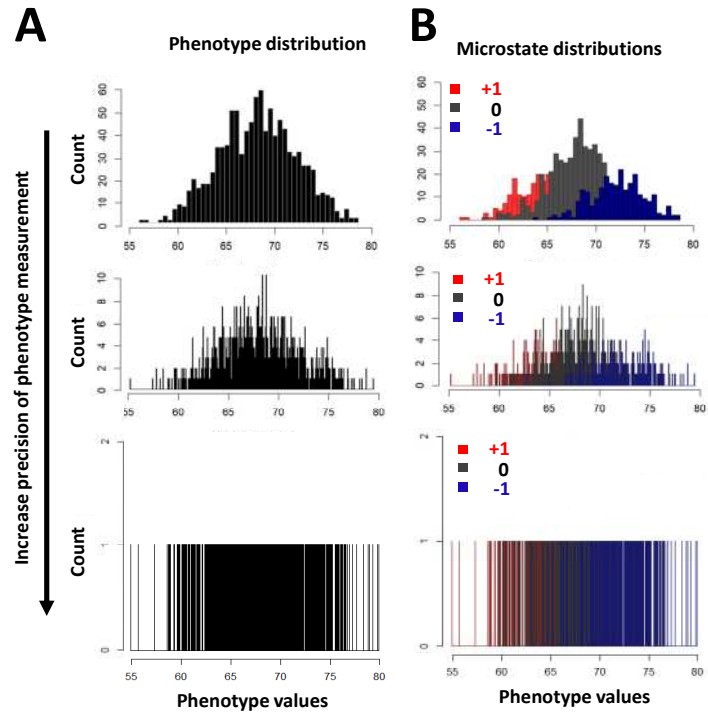

**Figure 1:** Traditional methods to map genotype to phenotype rely on probability density functions that are formed through the grouping of data into categories. The way GWAS proceed is by decomposing the phenotype distribution density function (A-top) onto genetic microstates (B-top). Such decomposition permits the analysis of averages and variances to map genotype to phenotype. The central issue however is that since categories are used more precise inferences can only come with, and are only legitimized by, a reduction in the width of categories. As a reduction in the width of categories implies a larger population size, an increase in the precision of inferences with GWAS is linked to an increase in the sample size. In order to overcome this issue one way of doing is to deconstruct density functions and wonder what would happen if one were able to reduce the width of categories without changing the sample sizes (A and B from top-to-bottom). The mathematical object that emerges then is a barcode and GIFT is a theory that allows a precise analysis of barcodes.

Let now us concentrate on a different SNP in the genome, noted SNP2, that is not associated with the phenotype. For SNP2 the configuration of microstates should appear as random, namely that the colour of bars should not be partitioned, or segregated, as shown in Fig.1B-bottom. An example of such case could be,

**SNP2: [-1, +1, 0, -1, -1, +1, +1, ..., -1, +1, +1, 0, -1, 0, +1, ..., 0, 0, -1, +1, 0, +1, -1]**     (2)

The difference between (1) and (2), namely the presence or absence of genotype-phenotype association, lies in the notion of 'scrambling' or 'mixing' of microstates. Naturally, SNP1 and SNP2 are different SNPs since they correspond to different genome positions and as a result, they are not directly comparable. However, one could artificially scramble the string corresponding to SNP1 and we would know that such a scrambled state is the configuration in which no association is possible between the genotype and the phenotype. Importantly, each string of microstates in the genome can be compared to its negative control through its scrambling. So, it is possible to visualise how genotype-phenotype association can be deduced with this simple method.

To conclude, one knows that increasing the precision in the measurement of phenotypic values leads us to consider barcodes as opposed to PDFs; and that the overall configuration of microstates is indicative of genotype-phenotype associations. Whilst interesting, those observations are still incomplete. Indeed, to use such observations one needs to: (i) find a way to generate them practically, i.e., experimentally, and; (ii) capture the different configurations using a theoretical framework that, not only agrees with current GWA method (Fisher theory), but generalises it.

### 1.2. Experimental (real-life) protocol to determine genotype-phenotype association using GIFT

Noting here that scrambling SNP1 above is equivalent to losing information on the phenotypic values. One can also say, conversely, that it is the information on phenotypic values that transforms the scrambled state into an ordered one. With this last remark one can envisage an experimental protocol to detect genotype-phenotype associations.

Let us consider a population of 'N' genotyped individuals, where the phenotype of interest has been measured precisely enough such that each individual has a unique phenotype value (as seen above Fig.1A and Fig.1B). Concentrating on a particular genome position for all individuals, if there are $N_{+1}$ genetic microstates of +1, $N_0$ genetic microstates of 0 and, $N_{-1}$ genetic microstates of -1, the genetic microstate frequencies, $\omega_+^0$, $\omega_0^0$ and $\omega_-^0$, for this particular genome position are determined by, $N_{+1}/N = \omega_+^0$, $N_0/N = \omega_0^0$ and $N_{-1}/N = \omega_-^0$. Now we consider two configurations, and to illustrate these two configurations, we assume the following setting as an example: The individuals are horses in a yard where adjacent to this yard there are two rows of N non nominative aligned paddocks each numbered $i = 1 \dots N$ (Figure 2A). The variable, $i$, shall be referred as the (paddock) position.

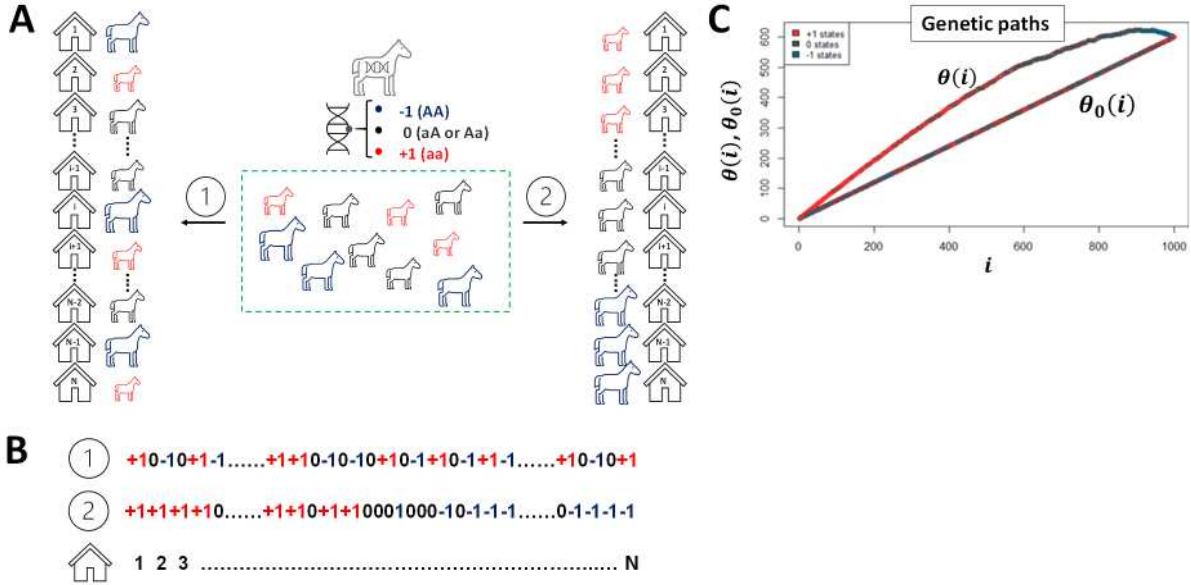

**Figure 2:** When a new method is promulgated it is essential to provide a way to use the method using a real setting. Let us consider a population of 'N' genotyped individuals that we shall assume are horses in a yard represented by the green-dashed rectangle in (A). Let us also consider that the phenotype of interest has been measured precisely enough such as each individual horse has a unique phenotype value. For diploid organisms and for a binary (bi-allelic, A or a) genetic marker, any microstate (genotype) can only take three values that we shall write as '+1' (in red), '0' (in black) and '-1' (in blue) corresponding to genotypes aa, Aa and AA, respectively. After genotyping individuals, one knows the genetic microstates frequencies. Concentrating on a particular genome position or a given single nucleotide polymorphism (SNP) in the genome for all individuals. If there are '$N_{+1}$' genetic microstates of '+1', '$N_0$' genetic microstates of '0' and '$N_{-1}$' genetic microstates of '-1', the genetic microstate frequencies for this particular genome position or SNP is $N_{+1}/N = \omega_+^0$, $N_0/N = \omega_0^0$ and $N_{-1}/N = \omega_-^0$. To develop the model further we consider two configurations and to illustrate those two configurations, we assume the following setting as shown in (A): The horses are in a yard where adjacent to this yard there are 'N' non nominative aligned paddocks numbered $i = 1 \dots N$. The variable 'i' shall be also referred as the (paddock) position. In the first configuration ①, the horses are allocated randomly to paddocks, that is, there is no information about any phenotype values. Once in the paddocks one concentrates on the SNP aforementioned and, starting from the first paddock, one notes the genetic microstate of individuals and corresponding paddock position. As a result, when individuals are randomly selected their genetic microstates taken together is a disordered string of '+1', '0' and '-1' as shown in (B) top-string. The cumulative sum of the genetic microstates found, is noted as '$\theta_0(i)$' where 'i' is the paddock's number or, equivalently, the position in the string of microstates. Given the random allocation, the probabilities of finding '+1', '0' or '-1' as genetic microstate at any position 'i' in the string (or in any paddock) are $\omega_+^0$, $\omega_0^0$ and $\omega_-^0$, respectively with a resulting cumulative sum that is: $\theta_0(i) = (+1 \cdot \omega_+^0 + 0 \cdot \omega_0^0 - 1 \cdot \omega_-^0)i$. '$\theta_0(i)$' and is therefore a straight line as shown in (C). We shall call '$\theta_0(i)$' the 'default genetic path'. In the second configuration ②, after selecting on phenotype the horses are ranked by phenotypic value. For example, if the phenotype is the height, the smallest horse is allocated to the first paddock and the highest horse the last one as shown in (A). The new cumulative sum of microstates is now calculated at each position in the microstate string. If an association exists between the genome position considered and the phenotype, then one may expect the genotypes aa and AA, i.e. '+1' and '-1', to be in the first and last positions, respectively; and the genotype Aa, i.e. '0', to be in the intermediate ones as shown in (B). As the same genome position has been considered between the two configurations, the total number of '+1', '0' and '-1' remains the same but what is different between the first and the second configurations, is the resulting shape of the cumulative sum of genetic microstates. In the second configuration the new cumulative sum noted '$\theta(i)$' and defined as the 'phenotype-responding genetic path' is not a straight line but a curve as shown in (C). Let us note $\omega_+(i)$, $\omega_0(i)$ and $\omega_-(i)$ the occurrence probabilities of the genetic microstates '+1', '0' and '-1' in this second configuration: $\theta(i) = \sum_1^i \left( +1 \cdot \omega_+(j) + 0 \cdot \omega_0(j) - 1 \cdot \omega_-(j) \right)$. As a result, the signature of a gene interacting with the phenotype when considering the two aforementioned genetic paths is the difference: '$\theta(i) - \theta_0(i)$', including a conservation relation since the two paths must meet when i = N, i.e. $\theta(N) = \theta_0(N)$ for they contain the same number of '+1', '0' and '-1' (see C). The difference '$\theta(i) - \theta_0(i)$' then describes how a variation in phenotype values is 'seen', or evolves, at the genetic level. In other words, '$\theta(i) - \theta_0(i)$' can be defined as the projection of the phenotype in the genetic space.

In the first configuration, see ① in Fig.2A, the horses are allocated randomly to paddocks; that is, the information on phenotype values is either absent or not used. Concentrating on genetic microstates found in paddocks, the random allocation is then equivalent to the scrambled state seen above corresponding to a disordered string of microstates, +1, 0 and -1 (Fig.2B). As in this case the probability of finding the genetic microstate +1, 0 or -1, at any position, i, is constant and given by, $\omega_+^0$, $\omega_0^0$ or $\omega_-^0$, respectively, where $\omega_+^0 + \omega_0^0 + \omega_-^0 = 1$; one can use the cumulative sum of the genetic microstates found to characterise the lack of genotype-phenotype associations. Let us, in this context, note by $\theta_0(i)$ such a cumulative sum where, i, is the paddock's number or, equivalently, the position in the string of microstates. The reason for using a cumulative sum in place of other more complicated formula resides in the fact that the microstate frequencies, $\omega_+^0$, $\omega_0^0$ and $\omega_-^0$, are constants. Indeed, in this case the cumulative sum of microstates can then be written, $\theta_0(i) = (+1 \cdot \omega_+^0 + 0 \cdot \omega_0^0 - 1 \cdot \omega_-^0)i$. As a conclusion, when genotype-phenotype associations are absent, $\theta_0(i)$, is a straight line. We call, $\theta_0(i)$, the default genetic path (Figures 2B and 2C).

In the second configuration, see ② in Fig.2A, after selecting the phenotype, the horses are ranked by their phenotypic value and sent to paddock accordingly. Note that as phenotypic values are distinct there is no possibility to send two horses in the same paddock. For example, if the phenotype is height, the smallest horse is allocated to the first paddock, and the highest horse is allocated to the last one. The new cumulative sum of the microstates is then calculated at each position in the microstate string. If an association exists between the genome position and the phenotype, then one may expect a change in the configuration of the string of microstates (Fig.2B), in turn impacting the new resulting cumulative sum of microstates. Thus, the only difference between the first and second configurations is the resulting shape of the cumulative sum of genetic microstates. Noting $\omega_+(i)$, $\omega_0(i)$ and $\omega_-(i)$ the occurrence probabilities of the genetic microstates +1, 0 and -1 in the second configuration the new cumulative sum is then: $\theta(i) = \sum_1^i\bigl(+1 \cdot \omega_+(j) + 0 \cdot \omega_0(j) - 1 \cdot \omega_-(j)\bigr)$; where $\theta(i)$ is defined as the 'phenotype-responding genetic path' (Fig.2C).

As a result, the signature of a gene interacting with the phenotype when considering the two aforementioned genetic paths can be thought as the difference: $\theta(i) - \theta_0(i)$, including a conservation relation since the two paths must meet when $i = N$, i.e., $\theta(N) = \theta_0(N)$, for they contain the same number of +1, 0, and -1 since the same SNP has been considered (Fig.2C). With this idea in mind and using physics terminology, one could say that '$\theta(i) - \theta_0(i) \neq 0$' because the phenotype acts as a 'informational field' to change the random configuration of microstates in the string. More precisely, because no information is available on the phenotype in the first configuration, the shape of the cumulative sum as a straight line arises from the random ordering of genetic microstates corresponding to the maximisation of the informational entropy of the system. When an association exists between genotype and phenotype the signature is a cumulative sum with a curved shape and in this case, the phenotypic 'field' competes with the entropy to order the string.

Posing the problem of genotype-phenotype associations this way, i.e., in term of phenotypic field, allows one to use the formalism as developed by physics to provide a mathematical expression for, $\omega_+(i)$, $\omega_0(i)$ and $\omega_-(i)$.

### 1.3. From genotype-phenotype associations to an analogy involving spin-magnetic field interactions: using physics to model GIFT

To comprehend how concepts from physics field theory can be used in genotype-phenotype mapping, one needs to think of the system of microstates in the string as being magnetic particles as in physics. Thus, let us consider a string composed of, $N_+$, $N_0$ and $N_-$ different and non-interacting magnetic particles, where $N_+ + N_0 + N_- = N$, each with a constant 'magnetic charge' (or spin), +1, 0 and -1, respectively. Also imagine that over time those non-interacting particles can exchange positions when no external constraint is applied onto the system. This 'hopping' (or diffusion) mechanism guaranties the disordered/scrambled configuration of microstates in the string when the system is at thermodynamic equilibrium without any external constraint (Fig.3A). Let now us assume that an external magnetic field is applied along the string as represented in Fig.3B.

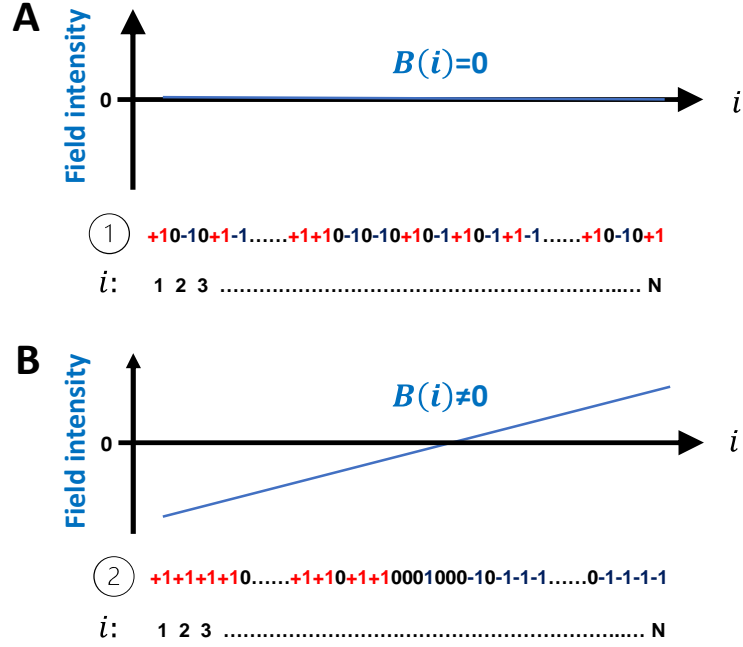

**Figure 3:** GIFT is a method that considers that the changes in the string's configuration between ① in (A) and ② in (B) arising upon knowledge/information acquisition on phenotype values, is linked or caused by a phenotypic field. In fact, GIFT is inspired from physics field theory. The notion of field can be understood as follow. Let us imagine that the microstates '1+', '0' and '-1' are similar to electric charges. When the field is null as shown in (A) the charges are not forced into a particular configuration and therefore the microstates are randomly allocated to positions. When the field is non-null as shown in (B) the charges will interact with the field and as a result the microstates will be forced to take a specific configuration. What matters with GIFT is the analytical shape and magnitude of the field.

We recall here that, in physics, the interaction between a magnetic field and a magnetic charge is proportional to the product between the field strength and the magnetic charge (or spin). Namely that the energy of a microstate of type, q, at the position, i, in the string is written as, $u_q(i) = q \times B(i)$, where, q, can either be, +1, 0 or -1. Using this formula one can then determine the optimal configuration of the string at the equilibrium when an external field is applied. To determine the configurations in Fig.3A and Fig.3B, one way to proceed is to minimise a functional corresponding to the equivalent of the total free energy of the system. This functional is composed the system's entropy, $-\sum_{i=1}^{N}\sum_{q\in\{+,0,-\}}\omega_q(i)\ln[\omega_q(i)]$, and energy, $\sum_{i=1}^{N}\sum_{q\in\{+,0,-\}}\omega_q(i)u_q(i)$. The free energy of the system in this case, $\mathcal{F}$, is written as,

$$\mathcal{F} = \sum_{i=1}^{N}\sum_{q\in\{+,0,-\}}\{\omega_q(i)u_q(i) - \omega_q(i)\ln[\omega_q(i)]\} \tag{3a}$$

One deduces then that, $\mathcal{F} \sim 0$, if, $\omega_q(i) \sim \exp[u_q(i)]$. This corresponds to the well-known Boltzmann's weight that is that the presence probability of a microstate of type, q, and at site, I, is function of the field, $u_q$, applied at that same position. However, the latter description is incomplete since three further constraints need to be considered including the total conservation of microstates,

$$\sum_{i=1}^{N} \sum_{q \in \{+, 0, -\}} \omega_q(i) = N \tag{3b}$$

the conservation of each microstate species in the string,

$$\sum_{i=1}^{N} \omega_q(i) = N_q \tag{3c}$$

as well as the normalisation of probability at each site in the string, i.e.,

$$\sum_{q \in \{+, 0, -\}} \omega_q(i) = 1 \tag{3d}$$

Comparing Eq.3d and Eq.3b, one sees that if Eq.3d is verified, then Eq.3b is also verified. Indeed, Eq.3b written as, $\sum_{i=1}^{N} \sum_{q \in \{+, 0, -\}} \omega_q(i) = N$, is also re-written, $\sum_{i=1}^{N} \omega_+(i) + \omega_0(i) + \omega_-(i) = \sum_{i=1}^{N} \omega_+(i) + (1 - \omega_+(i) - \omega_-(i)) + \omega_-(i) = \sum_{i=1}^{N} 1 = N$. As a result, one can expresses the free energy as a function of two microstates only and consequently a single constraint remains that is Eq.3c. In this context, one can use the Lagrange multipliers method to include the single remaining constraint in the functional, $\mathcal{H}$, representing the new free energy of the system, written under the form,

$$\mathcal{H} = \sum_{i=1}^{N} [\omega_+(i) u_+(i) + \omega_-(i) u_-(i) + (1 - \omega_+(i) - \omega_-(i)) u_0(i) - \omega_+(i) \ln(\omega_+(i)) - \omega_-(i) \ln(\omega_-(i)) - (1 - \omega_+(i) - \omega_-(i)) \ln(1 - \omega_+(i) - \omega_-(i))] + \lambda_+ [N_+ - \sum_{i=1}^{N} \omega_+(i)] + \lambda_- [N_- - \sum_{i=1}^{N} \omega_-(i)] + \lambda_0 [N - \sum_{i=1}^{N} (1 - \omega_+(i) - \omega_-(i))]$$

(4)

Where, $\lambda_p$ with $p \in \{+, 0, -\}$ are Lagrange multipliers. Finally, Euler-Lagrange method can be used to optimise the functional, $\mathcal{H}$, under the form, $\delta \mathcal{H} = (\partial_{\omega_+} \mathcal{H}) \delta \omega_+ + (\partial_{\omega_-} \mathcal{H}) \delta \omega_- = 0$, and where the partial derivatives must be nulls. Consequently, the conditions on each of the partial derivatives provide,

$$\ln\left(\frac{\omega_+(i)}{1 - \omega_+(i) - \omega_-(i)}\right) = u_+(i) - u_0(i) - \lambda_+ + \lambda_0 \tag{5a}$$

$$\ln\left(\frac{\omega_-(i)}{1-\omega_+(i)-\omega_-(i)}\right) = u_-(i) - u_0(i) - \lambda_- + \lambda_0 \tag{5b}$$

By using Eq.5a and Eq.5b together with Eq.3c it is then straight forward to determine,

$$\omega_q(i) = \frac{e^{u_q(i)-\lambda_q}}{Z(\{u_q(i)\})} \tag{6}$$

where, $Z(\{u_q(i)\}) = \sum_{q\in\{+,0,-\}} e^{u_q(i)-\lambda_q}$, is defined as the partition function. Given that the scrambled or random configuration, as seen in Fig.3A or given by ① in Fig.2A, is obtained when the fields are null, one deduces then that the probability to finding either microstate of type, p, is in this case, $N_q = \frac{e^{-\lambda_q}}{Z(\{0\})}$, where $Z(\{0\}) = \sum_{q\in\{+,0,-\}} e^{-\lambda_q}$ is the partition function when the fields are null. Finally, by replacing $e^{-\lambda_q}$ by $N_q Z(\{0\})$ where $q \in \{+, 0, -\}$ one can re-express Eq.6 as,

$$\omega_q(i) = \frac{\omega_q^0 e^{u_q(i)}}{\sum_{q\in\{+,0,-\}} \omega_q^0 e^{u_q(i)}} \tag{7}$$

With $\omega_q^0 = N_q/N$.

With Eq.7 the constraints given by Eq.3a and Eq.3c are fulfilled and Eq.3b can be re-written under a generic form as,

$$\sum_{i=1}^{N} \omega_p(i) = \sum_{i=1}^{N} \frac{\omega_p^0 e^{u_p(i)}}{\sum_{p\in\{+,0,-\}} \omega_p^0 e^{u_p(i)}} = N\omega_p^0 \tag{8}$$

As a result, this general formulation demonstrates that the probability of finding a microstate of type q in the string of microstates at any position is a function of the set of fields defined. While valid, this mathematical description departs from the physics intuition underscoring Fig.3 where a single field, noted B(i), was defined. However, a similar physics intuition can still be used as nothing impedes us to assume, boldly, that the fields must follow some specific symmetries (see below).

### 1.4. Omnigenic field and related phenotypic field symmetries.

Realistic GWAS have shown that with small effect sizes/small gene effects (which is the main area of concern of the current paper), dominance effects are often too small, and an additive model as suggested by Fisher works well enough [28]. As omnigenic traits are characterised by the involvement of many genes each with a small effect, the notion of dominance will be excluded from what will follow. As it turns out, very strong symmetries exist in Fisher's seminal papers (see Fig.1B). To postulate those symmetries, one needs to return in the space of phenotype values. We shall note by, $\widehat{U}_p(\Omega)$, the field applied on the microstate of type q as a function of the phenotype value, $\Omega$, found at the position, i, in the string. Based on Mendel observations Fisher's model assumes an anti-symmetrical segregation of microstates (see Fig.1B). It is possible to demonstrate in this context that for very small gene effects and when the dominance can be neglected, that the fields must verify, $\widehat{U}_+(\Omega) = -\widehat{U}_-(\Omega) = -\widehat{U}(\Omega)$, and, $\widehat{U}_0(\Omega) = 0$ [14], [15]. If, furthermore, one considers the normal distribution as a template for both microstates and phenotype and that the microstate variances are similar to the variance of the phenotype, then the fields are linear [14], [15]. If the microstate variances are different, then the fields are quadratic functions of the phenotype values. Those important results have been obtained through a coarse-graining process using the normal distribution as a template as Fisher did [14], [15]. Given that Fisher also assumed small effect sizes or small gene effects, this assumption is equivalent to considering that the magnitude of the phenotypic fields, i.e., its strength to rank microstates, is small. One can now turn back to the genetic path and determine the consequences linked to Fisher's theory (small gene/size effects).

**2. Genetic paths and conservation relation of genetic microstates.**

We recall that the difference in the genetic paths, $\theta(j) - \theta_0(j)$, can be written in the space of positions as (Fig.2C):

$$\theta(j) - \theta_0(j) = \left[\sum_{i=1}^{j} \omega_+(i) - \omega_-(i)\right] - \left[\sum_{i=1}^{j} \omega_+^0 - \omega_-^0\right] \qquad (9)$$

Let us then define, $di = (i + 1) - i$, and rewrite Eq.9 in the continuum limit as follow,

$$\theta(j) - \theta_0(j) = \left[\int_1^j (\omega_+(i) - \omega_-(i))di\right] - \left[\int_1^j (\omega_+^0 - \omega_-^0)di\right] \qquad (10)$$

It is worth recalling that as the number of microstates remains between the different genetic paths, one deduces $\theta(N) - \theta_0(N) = 0$. The latter relation is, by definition, the conservation relation of genetic microstates.

As the fields are deduced from using phenotype values, Eq.10 needs to be re-expressed in the space of phenotypic values. Defining $\lambda(\Omega)$ as the rate of change in phenotype values between two consecutive individuals when they are ranked using their phenotypic value, one can redefine the increment in the space of positions as, $di = d\Omega/\lambda(\Omega)$. Noting $\Omega_1$ and $\Omega_N$ the first and last phenotype values upon ranking the individuals one also deduces that, $\int_{\Omega_1}^{\Omega_N} \frac{d\Omega}{\lambda(\Omega)} = N - 1$. As a result, considering $\omega_+^0 + \omega_0^0 + \omega_-^0 = 1$, the conservation relation issued from Eq.10 is transformed as,

$$\hat{\theta}(\Omega_N) - \hat{\theta}_0(\Omega_N) = \left[\int_{\Omega_1}^{\Omega_N} \frac{\omega_+^0 e^{\widehat{U}_+(x)} - \omega_-^0 e^{\widehat{U}_-(x)}}{\omega_+^0 e^{\widehat{U}_+(x)} + \omega_0^0 e^{\widehat{U}_0(x)} + \omega_-^0 e^{\widehat{U}_-(x)}} \frac{dx}{\lambda(x)}\right] - \left[\int_{\Omega_1}^{\Omega_N} \frac{\omega_+^0 - \omega_-^0}{\omega_+^0 + \omega_0^0 + \omega_-^0} \frac{dx}{\lambda(x)}\right] = 0 \quad (11)$$

The 'hat' is added to note that one is now working in the space of phenotypic values. One notes that the second square bracket in the right-hand side of Eq.11 is similar to the first one provided that the fields are null. Setting for the fields, $2\widehat{U}_+(x) = \widehat{U}(x) + \Delta\widehat{U}(x)$ and $2\widehat{U}_-(x) = \widehat{U}(x) - \Delta\widehat{U}(x)$; and for the microstates frequencies, $2\omega_+^0 = \omega_0 + \Delta\omega_0$ and $2\omega_-^0 = \omega_0 - \Delta\omega_0$; one deduces that the condition linked to the conservation of genetic microstates given by, $\hat{\theta}(\Omega_N) - \hat{\theta}_0(\Omega_N) = 0$, can then be rewritten as,

$$\frac{1}{N-1} \int_{\Omega_1}^{\Omega_N} \frac{\Delta\omega_0 ch(\Delta\widehat{U}(\Omega)/2) + \omega_0 sh(\Delta\widehat{U}(\Omega)/2)}{(1-\omega_0) e^{\widehat{U}_0(\Omega) - \widehat{U}(\Omega)/2} + \Delta\omega_0 sh(\Delta\widehat{U}(\Omega)/2) + \omega_0 ch(\Delta\widehat{U}(\Omega)/2)} \frac{d\Omega}{\lambda(\Omega)} = \frac{\omega_+^0 - \omega_-^0}{\omega_+^0 + \omega_0^0 + \omega_-^0} \quad (12)$$

As $\Delta\omega_0/\omega_0 \in [-1; +1]$, one can also rewrite this ratio as a hyperbolic tangent under the form $\frac{\Delta\omega_0}{\omega_0} = \frac{sh(\varphi)}{ch(\varphi)} = th(\varphi)$. Let now us introduce Fisher's symmetry as published in his seminal paper [1] and assume that the dominance is null and the fields act anti-symmetrically on the homozygote microstates, namely, $\widehat{U}_+(\Omega) = -\widehat{U}_-(\Omega)$ and $\widehat{U}_0(\Omega) = 0$. Posing $\widehat{U}(\Omega) = \widehat{U}_+(\Omega) - \widehat{U}_-(\Omega)$, these assumptions transform Eq.12 as,

$$\frac{1}{N-1} \int_{\Omega_1}^{\Omega_N} \frac{sh(\widehat{U}(\Omega) + \varphi)}{\left(\frac{1-\omega_0}{\omega_0}\right) ch(\varphi) + ch(\widehat{U}(\Omega) + \varphi)} \frac{d\Omega}{\lambda(\Omega)} \sim \omega_0 th(\varphi) \quad (13)$$

One can then relate the meaning of the genetic constant, $\left(\frac{1-\omega_0}{\omega_0}\right) ch(\varphi)$, to Hardy-Weinberg equilibrium. Indeed, rewriting this constant as, $\left(\frac{1-\omega_0}{\omega_0}\right) ch(\varphi) = \left(\frac{1-\omega_0}{\omega_0}\right) \frac{1}{\sqrt{1-(th(\varphi))^2}} = \frac{1-\omega_0}{\sqrt{(\omega_0)^2 - (\Delta\omega_0)^2}}$. Given that, $\omega_0 = \omega_+^0 + \omega_-^0$, $\Delta\omega_0 = \omega_+^0 - \omega_-^0$, and $1 - \omega_0 = \omega_0^0$ one deduces that $\left(\frac{1-\omega_0}{\omega_0}\right) ch(\varphi) = \frac{1}{2} \frac{\omega_0^0}{\sqrt{\omega_+^0 \omega_-^0}}$. Let us assume then that the microstate frequencies are in Hardy-Weinberg ratio, namely $\omega_0^0 = 2pq$ and $\omega_+^0 \omega_-^0 = p^2 q^2$, where p and q=1-p are the allele frequencies, one deduces then, $\left(\frac{1-\omega_0}{\omega_0}\right)\text{ch}(\varphi) = 1$. Therefore, the parameter $\left(\frac{1-\omega_0}{\omega_0}\right)\text{ch}(\varphi)$ informs on how close from the Hardy-Weinberg ratio are the allele frequencies.

Setting a new integration variable, z, such as, $\Omega = z + \left(\frac{\Omega_N + \Omega_1}{2}\right)$; as $\Omega_N > \Omega_1$ one deduces then that Eq.13 is transformed to,

$$\int_{-\frac{\Delta\Omega}{2}}^{+\frac{\Delta\Omega}{2}} \frac{\text{sh}(\widetilde{U}(z)+\varphi)}{\left(\frac{1-\omega_0}{\omega_0}\right)\text{ch}(\varphi)+\text{ch}(\widetilde{U}(z)+\varphi)} \tilde{g}(z)dz \sim \omega_0 \text{th}(\varphi) \tag{14}$$

Where, $\Delta\Omega = \Omega_N - \Omega_1$, $\tilde{g}(z) = \langle\lambda\rangle/\lambda\left(z + \left(\frac{\Omega_N+\Omega_1}{2}\right)\right)$ with $\langle\lambda\rangle = 1/(N-1)$ and $\widetilde{U}(z) = \widehat{U}\left(z + \left(\frac{\Omega_N+\Omega_1}{2}\right)\right)$. Eq.14 can be solved only when $\widetilde{U}(z)$ and $\tilde{g}(z)$ are given. Recalling that $\tilde{g}(z)$ is a distant relative of the distribution density function of the phenotype, it is possible to rewrite Eq.14 differently.

To demonstrate that $\tilde{g}(z)$ is a distant relative of the distribution density function of the phenotype let us consider three consecutive phenotype values from the phenotype barcode that we shall note, $\Omega_i$, $\Omega_{i+1}$ and $\Omega_{i+2}$. As by definition, $\Omega_{i+1} - \Omega_i = \lambda(\Omega_i)$ and $\Omega_{i+2} - \Omega_{i+1} = \lambda(\Omega_{i+1}) = \lambda(\Omega_i + \lambda(\Omega_i))$, by adding these relations one deduces, $\Omega_{i+2} - \Omega_i = \lambda(\Omega_i) + \lambda(\Omega_i + \lambda(\Omega_i))$. Given that half of the interval of phenotypic values defined by, $\Omega_{i+2} - \Omega_i$, is the typical free phenotypic space available for the individual 'i + 1', the ratio given by $\sim 2/(\Omega_{i+2} - \Omega_i)$ is then similar to a concentration in the phenotypic space. Recall that in chemistry the concentration, C, is expressed as the number of molecules, N, contained in a volume, V, and is given by, $C = \frac{N}{V} = \frac{1}{V/N}$. Consequently, the parameter $V/N = v$ is the volume that one molecule occupies, and one can rewrite $C = 1/v$. Similarly, the function $h(\Omega)$ defined by, $h(\Omega) = \frac{2}{\lambda(\Omega)+\lambda(\Omega+\lambda(\Omega))}$, is analogous to the phenotypic space that one individual having the phenotype value $\Omega$, occupies. Let now us turn to the distribution density function of the phenotype. As the distribution density function of the phenotype defines the number of individuals, $\Delta N$, having a phenotype value comprised between $\Omega$ and $\Omega + \Delta\Omega$, one can define, $\Delta N = NP(\Omega)\Delta\Omega$, where $P(\Omega)$ is the distribution density function of the phenotype. One deduces then that the typical phenotypic space occupied by one individual in the category concerned is, $\frac{1}{\Delta\Omega/\Delta N} = NP(\Omega)$. Let now us assume that the population sampled is very large, one can rewrite, $h(\Omega) = \frac{2}{\lambda(\Omega)+\lambda(\Omega+\lambda(\Omega))} \sim \frac{2}{\lambda(\Omega)(2+\lambda'(\Omega))} \sim \frac{1}{\lambda(\Omega)} \sim NP(\Omega)$. By considering that a large population has been sampled, i.e., $N \gg 1$, one deduces therefore,

$$\tilde{g}(z) = \frac{\langle\lambda\rangle}{\lambda\left(z+\left(\frac{\Omega_N+\Omega_1}{2}\right)\right)} \sim \frac{N}{N-1} P\left(z+\left(\frac{\Omega_N+\Omega_1}{2}\right)\right) \sim \tilde{P}(z). \tag{15}$$

As the population sampled is very large and since $P(\Omega_1) \ll 1$ and $P(\Omega_N) \ll 1$, one can use the property of convergence of probability density functions to extend the integration integral. In this context, it is possible to rewrite Eq.14 as,

$$\int_{-\infty}^{+\infty} \frac{sh(\tilde{U}(z)+\varphi)}{\left(\frac{1-\omega_0}{\omega_0}\right)ch(\varphi)+ch(\tilde{U}(z)+\varphi)} \tilde{P}(z)dz \sim \omega_0 th(\varphi) \tag{16}$$

While extending the integration interval is mathematically valid given the convergence of probability density functions, any solution from Eq.16 needs to be benchmarked against the set of realistic phenotypic values, as not doing so can lead to unreal solutions concerning the genetic variables $\varphi$ and $\omega_0$.

### 3. Genotype-phenotype mapping, going beyond averages.

In his seminal paper Fisher used the normal distribution [1] and by defining the notion of gene effect, he gave a strong meaning to the notion of average to explain how genes in-form phenotypes. Said differently, for Fisher distribution density function of the phenotype and of microstates are paramount to define genotype-phenotype association. With GIFT however the viewpoint is different since genotype-phenotype association can be defined irrespectively of any distribution density functions. Indeed, a genotype-phenotype association can be inferred in the space of positions without involving any distribution density function (see Fig.2). In this context the phenotypic fields have been defined to explain why information on phenotypic values (their ranking) enacts on the positioning of microstates (see Figure 1 and the strings given by (1) and (2)) prefiguring, in turn, an association between the genotype and the phenotype. As GIFT is not concerned initially by the moments of the phenotype distribution density function the next question is: Can any phenotype distribution density function be used such as to exclude the mathematical definition of moments (e.g., mean or average) of the phenotype, while allowing Eq.14 to generate solutions?

For example, let us assume a linear phenotypic field and a phenotype distribution density function identical to Cauchy's distribution defined as, $\widetilde{P}(z) = \frac{1}{\pi}\frac{\gamma}{\gamma^2+z^2}$, where $\gamma$ is a scaling parameter. Cauchy's distribution is very distinctive as while the mode and median are well defined, it does not have a mean, variance, or higher moments. Said differently, if a set of phenotypic values were to follow Cauchy's distribution, no average, i.e., no gene effect, could be defined if Fisher theory, i.e., method of averages, were used. With GIFT however, Cauchy's distribution is not a problem. Indeed, since the method that defines GIFT dissociates the presence probability of microstates linked to a particular phenotype value from the distribution density function of that phenotype, whether or not an average exists is not meaningful. What is more is that a linear phenotypic field can also be postulated for reason of symmetry if need be. While a linear phenotypic field would agree with Fisher's theory, the absence of average would make it impossible for Fisher to extract information regarding genotype-phenotype association. The point that is emphasized here again is the potential generalisation that GIFT can provide in impossible cases as far as classic GWA methods are concerned.

Let now us assume the field be given by an affine function of the form $\widetilde{U}(z) = az + b$ coherent with Fisher's theory, where the parameter, a, is similar to the gene effect [14], [15]. Provided the distribution density function of the phenotype is known, the result from Eq.16 should be a function, G, involving the parameters a, b, $\varphi$ and $\omega_0$ written under the form, $G(a, b, \varphi, \omega_0) \sim 0$. That is to say that depending on the mathematical form of $G(a, b, \varphi, \omega_0)$, for a given values of a and b, different values of $\varphi$ and $\omega_0$ are in theory possible. Similarly, for a set of values $\varphi$ and $\omega_0$ different values for a a and b are possible.

Let now consider Cauchy's distribution together with the linear field above, Eq.16 can be rewritten as,

$$\int_{-\infty}^{+\infty} \frac{sh(az+b+\varphi)}{\left(\frac{1-\omega_0}{\omega_0}\right)ch(\varphi)+ch(az+b+\varphi)} \frac{1}{\pi}\frac{\gamma}{\gamma^2+z^2} dz \sim \omega_0 th(\varphi) \qquad (17)$$

This integral can be solved in the complex map using Cauchy's integral theorem. In this complex map specific symmetries arise and in particular one sees that Eq.17 is $2i\pi$-periodic in regard to $\varphi$, namely that replacing $\varphi$ by $\varphi \pm 2i\pi n$, where n is a relative integer, will not change Eq.17. Finally, posing $w = az + b + \varphi$ as a new variable and assuming it being a complex number, Eq.17 is transformed to,

$$\int_{-\infty}^{+\infty} F(w)H(w - b - \varphi)dw \sim \omega_0 \text{th}(\varphi) \tag{18a}$$

where,

$$H(w - b - \varphi) = \frac{1}{\pi}\frac{a\gamma}{(a\gamma)^2 + (w - b - \varphi)^2} \tag{18b}$$

and

$$F(w) = \frac{\text{sh}(w)}{\left(\frac{1-\omega_0}{\omega_0}\right)\text{ch}(\varphi) + \text{ch}(w)} \tag{18c}$$

Cauchy's integral theorem states that the integral in the left-hand side Eq.18a is equal to zero provided the contour integral chosen in the complex map excludes all poles, namely exclude the complex values of w for which the denominators of the integrand are null. The poles are determined by the equations,

$$(a\gamma)^2 + (w - b - \varphi)^2 \sim 0 \tag{19a}$$

$$\left(\frac{1-\omega_0}{\omega_0}\right)\text{ch}(\varphi) + \text{ch}(w) \sim 0 \tag{19b}$$

One deduces that for Eq.19a the solutions are always, $w_1 = b + \varphi + ia\gamma$ and $w_2 = \bar{w}_1 = b + \varphi - ia\gamma$, where $\bar{w}_1$ is the conjugated complex number of $w_1$ and i is the imaginary number ($i^2 = -1$). For Eq.19b however, the solutions are defined as a function of the numerical value of, $\left(\frac{1-\omega_0}{\omega_0}\right)\text{ch}(\varphi)$, and since, $\left(\frac{1-\omega_0}{\omega_0}\right)\text{ch}(\varphi) \geq 0$, two different cases need to be discussed.

Assume firstly that $\left(\frac{1-\omega_0}{\omega_0}\right)\text{ch}(\varphi) \geq 1$, one can then define a real value $w_0$ such as, $\left(\frac{1-\omega_0}{\omega_0}\right)\text{ch}(\varphi) = \text{ch}(w_0)$, and apply Cauchy integral theorem using the blue contour as shown in Fig.4A. In this case the poles of Eq.18c given by Eq.19b and included within the blue contour are given by, $w_0^+(n) = +w_0 + (2n + 1)i\pi$ and $w_0^-(n) = -w_0 + (2n + 1)i\pi$, where n is a relative integer number.

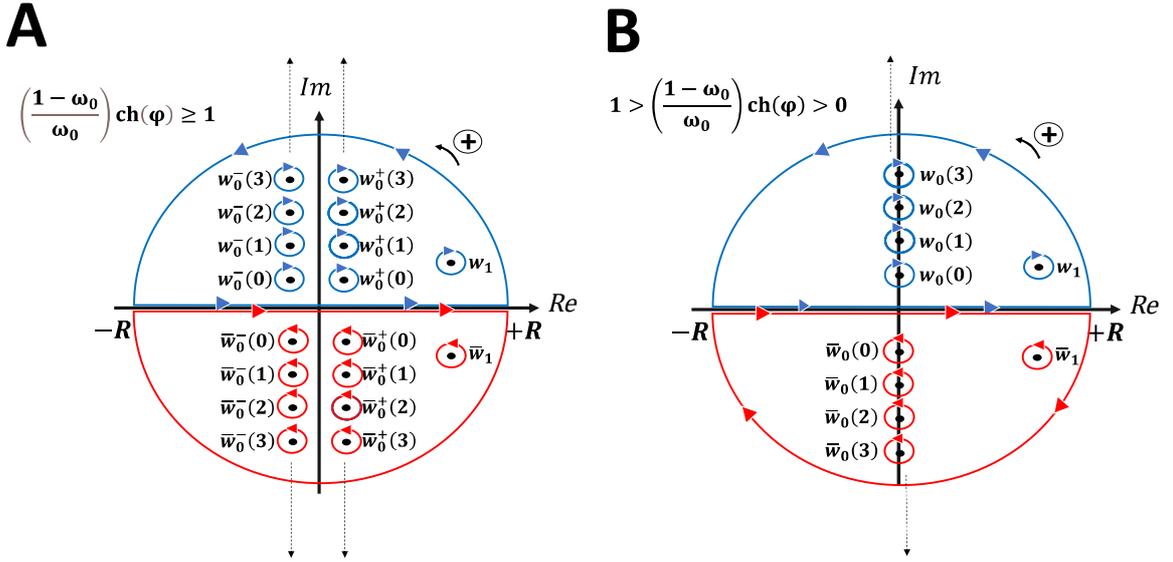

**Figure 4:** Contours used to apply Cauchy's integral theorem (see Appendices for details).

Cauchy's integral theorem together with the method of residues provides then (see Appendix B),

$$2i\pi \sum_{n=0}^{+\infty}[H(w_0^+(n) - b - \varphi) + H(w_0^-(n) - b - \varphi)] \sim \omega_0 \text{th}(\varphi) - F(w_1) \quad (20a)$$

Where,

$$H(w_0^+(n) - b - \varphi) + H(w_0^-(n) - b - \varphi) = \frac{a\gamma}{\pi}\left[\frac{1}{(a\gamma)^2 + (+w_0 + (2n+1)i\pi - b - \varphi)^2} + \frac{1}{(a\gamma)^2 + (-w_0 + (2n+1)i\pi - b - \varphi)^2}\right]$$

(20b)

And,

$$F(w_1) = \frac{\text{sh}(b + \varphi + ia\gamma)}{\text{ch}(w_0) + \text{ch}(b + \varphi + ia\gamma)} \quad (20c)$$

Eq.20a can be further simplified by observing that transforming, $\varphi \to \varphi - 2i\pi$, only impact on the left-hand side of Eq.20a since hyperbolic functions in the right-hand side are $2i\pi$-periodic. Recalling that $\text{ch}(w_0) = \left(\frac{1-\omega_0}{\omega_0}\right)\text{ch}(\varphi) = \left(\frac{1-\omega_0}{\omega_0}\right)\text{ch}(\varphi - 2i\pi)$ the transformation $\varphi \to \varphi - 2i\pi$ changes Eq.20a as,

$$2i\pi \sum_{n=0}^{+\infty}[H(w_0^+(n+1) - b - \varphi) + H(w_0^-(n+1) - b - \varphi)] \sim \omega_0 th(\varphi) - F(w_1) \quad (20d)$$

One can now change the initial value of, n, used for the summation to rewrite Eq.20d as,

$$2i\pi \sum_{n=1}^{+\infty}[H(w_0^+(n) - b - \varphi) + H(w_0^-(n) - b - \varphi)] \sim \omega_0 th(\varphi) - F(w_1) \quad (20e)$$

Subtracting Eq.20e from Eq.20a implies finally,

$$H(w_0^+(0) - b - \varphi) + H(w_0^-(0) - b - \varphi) \sim 0 \quad (20f)$$

Eq.20f has been obtained considering the transformation, $\varphi \to \varphi - 2i\pi$. However, as Eq.17 is $2i\pi$-periodic in $\varphi$ one could have considered also $\varphi \to \varphi - 4i\pi$ and through a similar reasoning as above deduce,

$$H(w_0^+(0) - b - \varphi) + H(w_0^-(0) - b - \varphi) + H(w_0^+(1) - b - \varphi) + H(w_0^-(1) - b - \varphi) \sim 0$$

$$(20g)$$

Given Eq.20f it follows then, $H(w_0^+(1) - b - \varphi) + H(w_0^-(1) - b - \varphi) \sim 0$. Consequently, through recurrence one can demonstrate that the left-hand side of Eq.20e is always null and as a consequence,

$$\omega_0 th(\varphi) - F(w_1) \sim 0 \quad (20h)$$

Using now the red contour as represented in Fig.4A, one deduces a relation similar to Eq.20a given by (see Appendix B),

$$-2i\pi \sum_{n=0}^{+\infty}[H(\bar{w}_0^+(n) - b - \varphi) + H(\bar{w}_0^-(n) - b - \varphi)] \sim \omega_0 th(\varphi) - F(\bar{w}_1) \quad (20i)$$

With $\bar{w}_0^+(n) = +w_0 - (2n+1)i\pi$ and $\bar{w}_0^-(n) = -w_0 - (2n+1)i\pi$. Using the transformation $\varphi \to \varphi + 2i\pi$ together with a reasoning based on recurrence similar to the one performed above one deduces,

$$\omega_0 th(\varphi) - F(\bar{w}_1) \sim 0 \quad (20j)$$

As a final result concerning the condition $ch(w_0) = \left(\frac{1-\omega_0}{\omega_0}\right) ch(\varphi) \geq 1$ one finds a fundamental relation related to the conservation of genetic microstates given by, $F(\bar{w}_1) = F(w_1)$.

Assume now that $1 > \left(\frac{1-\omega_0}{\omega_0}\right) ch(\varphi) > 0$, one can then define an imaginary number $iw_0$ such as, $\left(\frac{1-\omega_0}{\omega_0}\right) ch(\varphi) = ch(iw_0) = cos(w_0)$, and apply Cauchy integral theorem using the blue and red contours drawn in Fig.4B. In this case the poles of Eq.18b given by Eq.19a are still $w_1 = \varphi + ia\gamma$ and $w_2 = \overline{w}_1 = \varphi - ia\gamma$, while the poles of Eq.18c given by Eq.19b and are now given by, $w_0(n) = +iw_0 + (2n+1)i\pi$ or $\overline{w}_0(n) = -iw_0 - (2n+1)i\pi$ when included in the blue or red contour, respectively. Let us concentrate on the blue contour firstly. Cauchy's integral theorem together with the method of residues provides then (see Appendix C),

$$2i\pi \sum_{n=0}^{+\infty} H(w_0(n) - b - \varphi) \sim \omega_0 th(\varphi) - F(w_1) \qquad (21a)$$

As the right-hand side is invariant through the transformation $\varphi \to \varphi - 2i\pi$ since hyperbolic functions are $2i\pi$-periodic. Imposing $\varphi \to \varphi - 2i\pi$ transforms Eq.21a as,

$$2i\pi \sum_{n=1}^{+\infty} H(w_0(n) - b - \varphi) \sim \omega_0 th(\varphi) - F(w_1) \qquad (21b)$$

As a result of subtracting Eq.21b from Eq.21a one deduces the condition,

$$H(w_0(0) - b - \varphi) \sim 0 \qquad (21c)$$

As performed above by using the transformation $\varphi \to \varphi - 4i\pi$, one could have similarly demonstrated by recurrence that,

$$\omega_0 th(\varphi) \sim F(w_1) \qquad (21d)$$

Concentrating now on the red contour, Cauchy's integral theorem together with the method of residues provides (see Appendix C),

$$2i\pi \sum_{n=0}^{+\infty} H(\overline{w}_0(n) - b - \varphi) \sim \omega_0 th(\varphi) - F(\overline{w}_1) \qquad (21e)$$

Let now use the property of invariance of the terms in the right-hand side member of Eq.21c, namely consider the transformation, $\varphi \to \varphi + 2i\pi$; one can then demonstrate that Eq.21d is transformed to,

$$2i\pi \sum_{n=1}^{+\infty} H(\overline{w}_0(n) - b - \varphi) \sim \omega_0 th(\varphi) - F(\overline{w}_1) \qquad (21f)$$

As a result, subtracting Eq.21f form Eq.21e imposes the condition,

$$H(\overline{w}_0(0) - b - \varphi) \sim 0 \qquad (21g)$$

Using the same reasoning as above one could have applied the transformation $\varphi \to \varphi - 4i\pi$, and by recurrence demonstrate that,

$$\omega_0 \text{th}(\varphi) \sim F(\overline{w}_1) \tag{21h}$$

As a final result concerning the condition, $1 > \text{ch}(iw_0) = \left(\frac{1-\omega_0}{\omega_0}\right)\text{ch}(\varphi) > 0$, one deduces a conservation relation concerning the genetic microstates given by, $F(\overline{w}_1) = F(w_1)$. Note that this relation is the same as the one found for the condition $\left(\frac{1-\omega_0}{\omega_0}\right)\text{ch}(\varphi) \geq 1$.

Using Eq.18c and Eq.20c to develop the imaginary and real parts of $F(\overline{w}_1) = F(w_1)$ one deduces finally a relation concerning the conservation of genetic microstates given by, $\sin(a\gamma)\left[\cos(a\gamma) - \left(\frac{1-\omega_0}{\omega_0}\right)\text{ch}(\varphi)\text{ch}(b+\varphi)\right] \sim 0$. As a result, Eq.22 is null either when,

$$\sin(a\gamma) \sim 0 \Rightarrow a\gamma \sim n\pi \tag{22a}$$

where n is a relative number, or when,

$$\cos(a\gamma) \sim \left(\frac{1-\omega_0}{\omega_0}\right)\text{ch}(\varphi)\text{ch}(b+\varphi) \tag{22b}$$

Note that as $\cos(a\gamma) \leq 1$ in Eq.22b, the only valid solutions for the microstates when Cauchy's distribution is used are those respecting also the condition: $(1 - \omega_0)\text{ch}(\varphi)\text{ch}(b+\varphi) \leq \omega_0$.

Eq.22a and Eq.22b are the result of Eq.17 deduced from the conservation relation given by Eq.10. Eq.10 is a meaningful relation since it encapsulates the fact that the ordered, i.e., $\hat{\theta}(\Omega)$, and random, i.e., $\hat{\theta}_0(\Omega)$, paths intersect in due course, i.e., $\hat{\theta}(\Omega_N) - \hat{\theta}_0(\Omega_N) = 0$. While the later relation is a conservation relation for the path, such conservation relation is only possible because the microstate frequencies are conserved. This suggests that further conservation relations can be determined.

### 4. The other conservation relation for genetic microstates.

As there are three different microstate frequencies and that their sum must always be equal to one, two conservation relations are sufficient. As $\hat{\theta}(\Omega)$ refers to the difference between the microstate frequencies '+' and '-', one can now use their sum to provide a further integral conservation relation, namely

$$\int_{-\infty}^{+\infty} \frac{\text{ch}(az+b+\varphi)}{\left(\frac{1-\omega_0}{\omega_0}\right)\text{ch}(\varphi)+\text{ch}(az+b+\varphi)} \frac{1}{\pi} \frac{\gamma}{\gamma^2+z^2} dz \sim \omega_0 \tag{24}$$

As done above, Eq.24 can be rewritten as,

$$\int_{-\infty}^{+\infty} \tilde{F}(w)H(w-b-\varphi)dw \sim \omega_0 \qquad (25a)$$

where,

$$H(w-b-\varphi) = \frac{1}{\pi} \frac{a\gamma}{(a\gamma)^2+(w-b-\varphi)^2} \qquad (25b)$$

and

$$\tilde{F}(w) = \frac{ch(w)}{\left(\frac{1-\omega_0}{\omega_0}\right)ch(\varphi)+ch(w)} \qquad (25c)$$

As seen above, solution will emerge as a function of the value that the parameter, $\left(\frac{1-\omega_0}{\omega_0}\right)ch(\varphi)$, takes.

Let now us assume that $\left(\frac{1-\omega_0}{\omega_0}\right)ch(\varphi) \geq 1$, Eq.24 can be solved using Cauchy's integral theorem using the two different contours as shown in Fig.4A (Appendix D). One deduces then,

$$+2i\pi \frac{1}{th(+w_0)} \sum_{n=0}^{+\infty}[H(w_0^+(n)-b-\varphi) - H(w_0^-(n)-b-\varphi)] \sim \omega_0 - \tilde{F}(w_1) \qquad (26a)$$

$$-2i\pi \frac{1}{th(+w_0)} \sum_{n=0}^{+\infty}[H(\overline{w}_0^+(n)-b-\varphi) - H(\overline{w}_0^-(n)-b-\varphi)] \sim \omega_0 - \tilde{F}(\overline{w}_1) \qquad (26b)$$

Using the recurrence relations to deduce Eq.20h and Eq.20j, one can deduce similarly that the left-hand sides of Eq.26a and Eq.26b are null and that Eq.26a and Eq.26b are valid provided: $\tilde{F}(w_1) \sim \tilde{F}(\overline{w}_1)$.

Let now us assume that, $1 > \left(\frac{1-\omega_0}{\omega_0}\right)ch(\varphi) > 0$, Eq.24 can be solved using Cauchy's integral theorem using the two different contours as shown in Fig.4B (Appendix E). One deduces then,

$$\frac{2i\pi}{th(iw_0)} \sum_{n=0}^{\infty} H(w_0(n)-b-\varphi) \sim \omega_0 - F(w_1) \qquad (27a)$$

$$\frac{2i\pi}{th(iw_0)} \sum_{n=0}^{\infty} H(\overline{w}_0(n)-b-\varphi) \sim \omega_0 - F(\overline{w}_1) \qquad (27b)$$

Using once again the recurrence relations to deduce Eq.21d and Eq.21h, one can deduce also that the left-hand sides of Eq.27a and Eq.27b are null and that Eq.27a and Eq.27b are valid provided, $\tilde{F}(w_1) \sim \tilde{F}(\overline{w}_1)$.

As a result, the new condition arising from $\tilde{F}(w_1) \sim \tilde{F}(\overline{w}_1)$ is then, $\text{sh}(b+\varphi)\sin(a\gamma) = 0$, which admits two solutions given by,

$$\sin(a\gamma) \sim 0 \Rightarrow a\gamma \sim n\pi \tag{28a}$$

or

$$b + \varphi \sim 0 \tag{28b}$$

## 5. Determination of genetic microstate frequencies in the case of a Cauchy-distributed phenotype: Case of homozygotes excess.

From the analyses carried out two set of solution emerge. The first set is given by Eq.22a or Eq.28a; and the second set by Eq.22b and Eq.28b. As the first set of do not involve microstate frequencies, this suggests in turn that the solutions have no real-life existence. One the other Eq.22b and Eq.28b are valid in a context where linear fields are used in agreement with Fisher's seminal idea following which the genetic microstates must be clearly segregated asymmetrically when genotype and phenotype are associated, the subset of gene microstates involved for a Cauchy-distributed phenotype can be defined as, i.e.,

$$\cos(a\gamma) \sim \left(\frac{1-\omega_0}{\omega_0}\right) \text{ch}(\varphi) \sim \frac{1}{2} \frac{\omega_0^0}{\sqrt{\omega_+^0 \omega_-^0}} \tag{29}$$

Since $\cos(a\gamma) \leq 1$, Eq.29 is only valid if $\omega_+^0 \omega_-^0 \geq (\omega_0^0)^2/4$. Recall that the last right-hand side term is a parameter linked to Hardy-Weinberg coefficient, see paragraph below Eq.13. Accordingly, solution to Eq.29 exists only if the proportion of homozygotes is in excess. Further assuming small gene effects Eq.29 becomes: $(a\gamma)^2 \sim 2 - \omega_0^0/\sqrt{\omega_+^0 \omega_-^0}$.

To conclude, Eq.29 demonstrates that thanks to GIFT genotype and phenotype can be associated without involving any averages.

**Discussion**

Identifying the association between phenotypes and genotypes is the fundamental basis of genetic analyses. Beginning with Mendel's work at the end of the 19$^{th}$ century, genotypes were inferred by tracking the inheritance of phenotypes between individuals with known relationships (linkage analysis). In recent years, the development of molecular tools, culminating in high-density genotyping and whole genome sequencing, has enabled DNA variants (e.g., single nucleotide polymorphisms (SNPs)) to be directly identified and phenotypes to be associated with genotypes in large populations of unrelated individuals through association mapping. However, while new biotechnologies have allowed us to probe DNA variants more accurately, the statistical tools and conceptual frameworks used to analyse data are still derived from a method established by Fisher more than a century ago. While Fisher method works very well when the gene effects are large, the method becomes limiting when small gene effects are involved. This limit manifest itself via the necessity to use unrealistically large data sets to prove associations. Here we show that the main cause to this failure is conceptual and that it stems from the postulate that random sampling of alleles at a gene would produce a continuous and normally distributed phenotype, requiring *de facto* the use of averages and variances.

As already alluded in the introduction, one reason for large datasets is linked to the utilisation of distribution density functions. Based on frequentist probability, distribution density functions are generated 'experimentally' using bar charts, namely by grouping data into bins or categories. However binning data results in loosing information since it is not possible to differentiate/distinguish data from within a given bin/category. Said differently, replacing individual data values by an average result in the loss of individual information. Additionally, as the width of bins/categories is linked to the notion of precision in phenotypic measurements, increasing precision levels to get access to very small gene effects implies reducing the width of bins/categories meaning, in turn, recruiting ever-larger populations to refine the distribution density functions and related inferences. This is why GWASs are data-consuming. Indeed, full precision can only occur when the population measured is infinite to reduce to zero the width of bins/categories. However, since an infinite population does not exist, a deal is agreed by fitting the bar charts using a continuous curve assuming, implicitly, the validity of distribution density functions in the continuum (asymptotic) limit. Doing so gives the impression that statistical parameters, e.g., average, variance or any other moments, possibly extracted from considering the fit in the continuum (asymptotic) limit have fundamental meanings. However, this impression is misleading since it is the continuum limit that is an approximation of the bar-chart, and not the converse. Consequently, as it is not possible to differentiate/distinguish data from within a given bin or category, using average and variance as ontological parameters describing bar charts in the continuum (asymptotic) limit means also that average and variance are defined through an assumed lack of information limiting, in turn, our understanding of living systems. To conclude, unless the continuum (asymptotic) limit can be justified, average, variance and the other moments of density functions deduced from using the continuum (asymptotic) limit are restricted in the information they can provide.

Usually, the latter statement is never formulated nor considered since the central limit theorem is called-in to justify: (i) the use of the normal distribution in the continuum (asymptotic) limit, (ii) the meanings of averages and variances at the population level and, (iii) the method used in GWASs, i.e., Fisher's method a.k.a. method of averages. The problem however is that the central limit theorem that gives rise to the normal distribution is based on specific assumptions that have been formulated in physics. Consequently, these specific assumptions are not always present in biological systems. Let us re-clarify this point. Previously known as the Law of Errors, the normal distribution was conceived to capture measurement or observational errors in physics. Unlike Biology driven by Evolution, physics is determined by the intemporal Laws of Nature. Causal determinism in physics is then central to capture measurable observables. Accordingly, the average conceptualised as a sort of stability resulting from the presence of deterministic Laws is a fundamental parameter in physics. Physics can also use the normal distribution resulting from the categorisation/binning of data to extract averages and variances because its constitutive elements are always the same when a given experiment is performed several times. As an example, let us assume that one wants to determine the mass of an electron through a series of repeated experiments. As Physics postulates that the mass of electrons from any atom are identical with regard to their mass, the binning or categorisation of data linked to the property of the mass of electrons measured does not lead to a loss of information. These arguments justify the use of frequentist probabilities. Finally, distribution density functions such as the normal distribution obtained mathematically in the continuum (asymptotic) limit can be used in physics since the Laws of Nature are intemporal. Consequently, any physics experiment can be repeated any time, possibly an infinite number of times over time.

To conclude, if one had to re-transcribe the set of physics assumptions in the field of Biology to legitimate the use of the normal distribution and give meanings to the average and variance, only an infinite population of un-evolving clones should be considered as indeed, a normal distribution arises as a result of measuring many times the same thing in identical circumstances. Therefore, the fact that in most case biological phenotypes are not perfectly normally distributed is central to demonstrate the uniqueness of biology, i.e., biology is not physics, also confirming the fundamental roles of diversity and evolution.

One may argue that non-parametric statistics can be used to generate inferences without involving the normal distribution. Indeed, non-parametric statistics can be used for biological phenotypes that are not normally distributed. While this point is perfectly valid, it raises a fundamental question regarding the meanings of the average and variance in these cases. In physics any average was meant to be measured, i.e., is necessarily meaningful since it is the consequence of the Intemporal Laws of Nature. Such affirmation is difficult to uphold in biology for all the reasons mentioned above. Thus, extracting an average and a variance from non-normally distributed data raise the question about their meaning.

When R.A. Fisher developed his theory, he was inspired by physics [29], a point that is particularly visible in his second paper entitled 'On the dominance ratio' published in 1923 when he writes: 'The distribution of the frequency ratio for different factors may be calculated from the condition that this distribution is stable, as is that of velocities in the Theory of Gases' [2]. However, assimilating biology to physics to extract averages lead to problem that are visible today under the form of incremental time/cost needed to run ever-more precise GWASs. While citing physics may sound convincing, such justification implies also that the biological object studied be conceptualised in a specific way, i.e., in a physics way. Since biology is not physics, this type of argument leads, in due course, to conceptual/epistemological issues that have already been noted above (c.f., 'un-evolving clones' or missing information) and are discussed in details elsewhere as far as genotype-phenotype mapping is considered [14].

It is for all the reasons aforementioned above that GIFT was formulated as an attempt to avoid the numerous pitfalls linked to the blind utilisation of the normal distribution.

With regard to the meaning of GIFT involving Cauchy-distributed phenotypes, it is remarkable that some debates we thought closed can re-emerge centuries later. In this context what happened in 1735 is very informative and looking closer at the initial proof of the Law of Errors, i.e., the normal distribution, produced by C.F. Gauss we can understand Cauchy's reserves and why such mathematical proof is labelled 'circular reasoning' [30]. One leaves the reader to read the proof in [13].

Finally, linking Cauchy's distribution to GIFT is significant in the field of genetics for two reasons. The first reason lies in the fact that GIFT is, by construction, a method that does not use the average or variance to determine how genotype and phenotype are associated. Using Cauchy's distribution together with GIFT was therefore important to demonstrate that genotype-phenotype mapping solutions are possible even in the case where average and variance cannot be defined. Note again that classic genotype-phenotype association methods would reject Cauchy's distribution as being a pathological distribution, i.e., an impossible case to treat. The second nonetheless important point concerns the notion of evolutionary rescue (reviewed in [31]). Population experiencing severe and abrupt stress can only avoid extinction provided adaptation happens. Adaptation is facilitated through the involvement of phenotypes beyond average ones, a.k.a. rare variants. Given that the tails of Cauchy's distribution are 'thicker' than the tails of the normal distribution, rare variant are rarer when the normal distribution is used and limited by the bulk of the distribution a.k.a. variance, when compared to the Cauchy distribution. In this context, finding a genetic basis for Cauchy-distributed phenotypes is important as large phenotypic fluctuations that arise may prime species to evolution as opposed to extinction.

**Conclusion**

This work confirms that average and variance are not required to map genotype to phenotype. In addition, this work provides the proof of principle that specific genotypes might be used for rapid evolution.

**Acknowledgements:** The author would like to thank Jonathan Wattis for fruitful discussions.

**Fundings:** The work was funded by Waltham and the University of Nottingham.

**Conflict of interest:** The author declares no conflict of interest.

**Appendix A: Differences and similarities between the normal and Cauchy distributions.**

The Normal distribution and the Cauchy distribution a.k.a. probability density functions (PDFs) are both probability distributions used in statistics to explain random variables having an infinite number of possible values within a defined range. While they have certain shared traits, they also exhibit unique features. The normal distribution, sometimes referred as the Gaussian distribution or the bell-shape curve, is one of the most popular probability distributions. Its symmetric bell-shaped curve is often considered as its key characteristic. The mean ($\mu$) and the standard deviation ($\sigma$) of the normal distribution are the two parameters that completely define the distribution. The shape of the curve is formulated by these parameters, with the mean representing the central tendency and the standard deviation regulating the spread or variability of data its PDF is given by: $P(x) = \frac{1}{\sigma\sqrt{2\pi}} \exp\left\{-\frac{1}{2}\left(\frac{x-\mu}{\sigma}\right)^2\right\}$, $\sigma^2 > 0$.

On the other hand, the Cauchy distribution is a continuous probability distribution with PDF given by,

$$P(x) = \frac{1}{\pi}\left[\frac{\gamma}{(x-x_0)^2 + \gamma^2}\right], \gamma > 0 \qquad (Eq.A1)$$

that looks like a normal distribution. Even though the resemblance exists, it represents a 'thinner' peak than the normal distribution does, and a different shape with heavy side tails increasing the likelihood of getting extreme values (see Fig.A1). Due to its large tails, the Cauchy distribution does not have finite moments. In particular, the mean and variance are undefinable. This means that gathering thousands of data points will not result in a more precise estimate of the mean and standard deviation than a single point. Hence, the Cauchy distribution is fully described, as Eq.A1 suggests, by the location parameter ($x_0$), which indicates the location of the peak, and the scale parameter $\gamma > 0$. As a result, the Cauchy distribution is famous for the fact that the expected value and standard deviation do not exist. To investigate this statement, a sample of a hundred variables, length of 1000, has been generated to visualise the differences between the expected mean and standard deviation in both cases (Normal and Cauchy). A plot as a function of the sample number is given in Fig.A2. Fig.A2 illustrates the lack of stability regarding the mean and standard deviation when Cauchy distribution is considered. In summary, the normal distribution is symmetric and has finite moments, while the Cauchy distribution is symmetric it lacks finite moments.

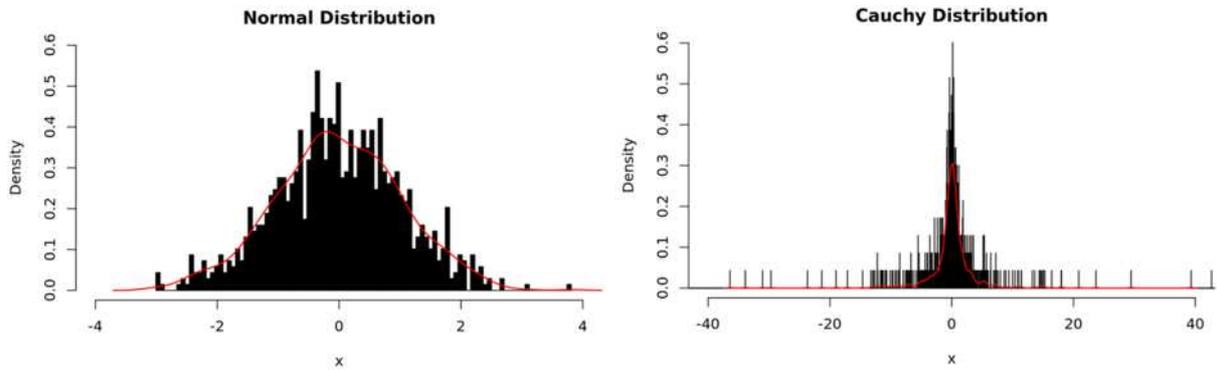

**Figure A1:** On the left, the normal distribution with mean $\mu = 0$ and standard deviation $\sigma = 1$. On the right, the Cauchy distribution with location $x_0 = 0$ and scale parameter $\gamma = 1$. Note the difference in the spreading of the random variable labelled 'x' (x-axis).

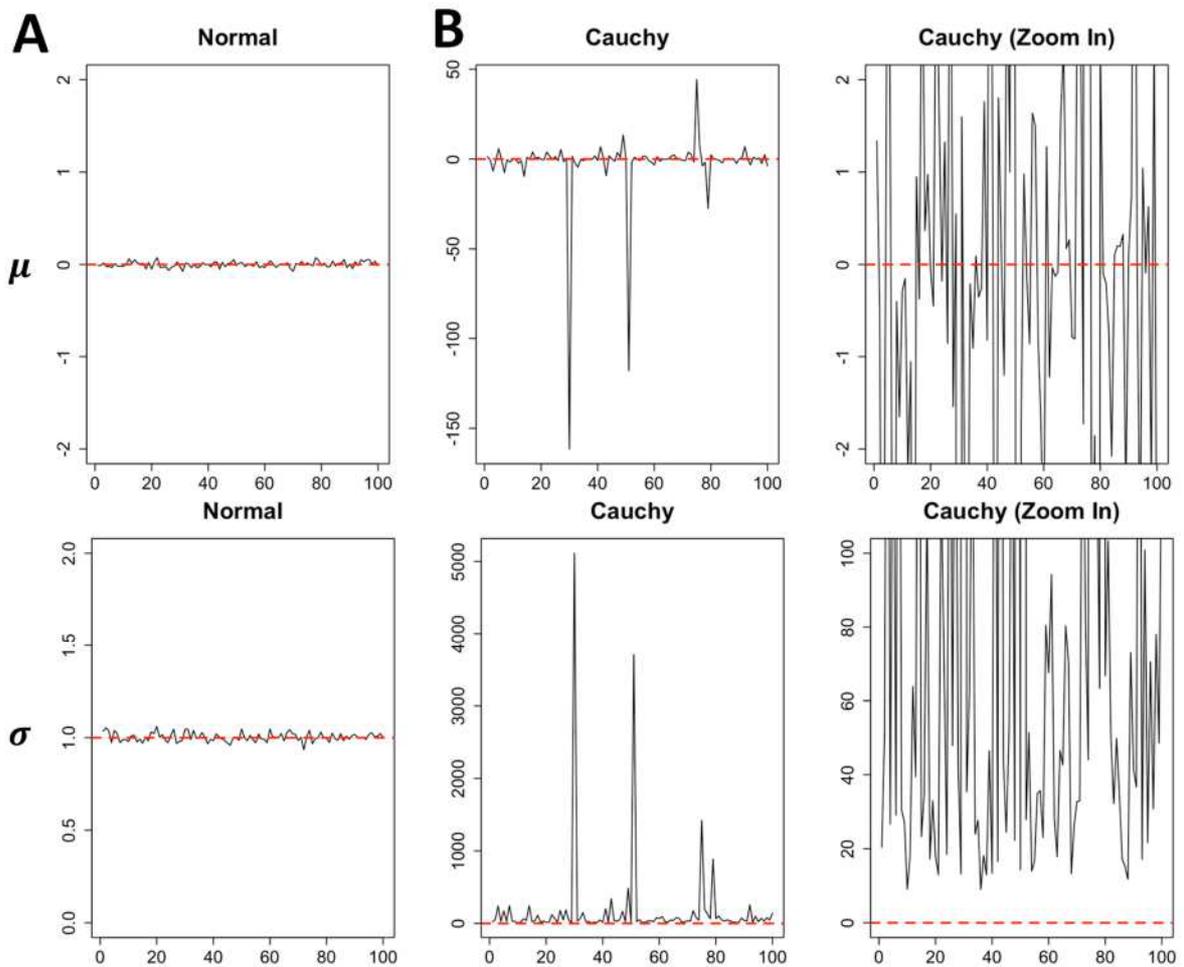

**Figure A2:** **(A)** Plots of expected mean (top panel) and standard deviation (bottom panel) of standard Normal distribution $N(0,1)$ and standard Cauchy distribution $C(0,1)$. Estimated mean of $N(0,1)$ is around zero and its standard deviation around one. In contrast, the estimated mean

and standard deviation of $C(0,1)$ fluctuate and are therefore not stable. **(B)** Magnified view of the mean (top) and standard deviation (bottom) of Cauchy distribution.

## Appendix B: Cauchy's Integral Theorem applied to genetic when: $\left(\frac{1-\omega_0}{\omega_0}\right)\text{ch}(\varphi) = \text{ch}(w_0) \geq 1$.

The objective is to use Cauchy Integral Theorem in the complex map to resolve the integral given by,

$$I = \int_{-\infty}^{+\infty} \frac{\text{sh}(w)}{\text{ch}(w_0)+\text{ch}(w)} \frac{a\gamma}{\pi} \frac{1}{(a\gamma)^2+(w-b-\varphi)^2} dw \tag{Eq.B1}$$

To remain coherent with the definitions from the main text, we shall rewrite the integral as,

$$I = \int_{-\infty}^{+\infty} F(w)H(w-b-\varphi)dw \tag{Eq.B2}$$

Integrals in the complex map can use continuous closed contours as opposed to integration intervals and Cauchy Integral Theorem stipulates that provided the continuous contour is drawn in such a way to exclude the poles of the integrand, then the integral is null on the continuous contour chosen. Consequently, assuming now $w$ is a complex number and considering a continuous closed contour noted $C$ in the complex map the integral one aims to resolve is,

$$J = \oint_C F(w)H(w-b-\varphi)dw \tag{Eq.B3}$$

Note that Cauchy Integral Theorem stipulates that if the closed contour excludes the poles, then, $J = 0$. As $\left(\frac{1-\omega_0}{\omega_0}\right)\text{ch}(\varphi) = \text{ch}(w_0) \geq 1$, the poles are represented as black dots in Fig.4A and depending on the sign of the poles' imaginary part, two contours can be draw that are represented by the blue and red contours.

Let us concentrate on the blue contour from Fig.4A. This continuous blue contour is the sum of a set of integrable elements including the large semi-circle, the diameter of the semi-circle as well as the small blue contours surrounding the poles. Note that as the overall contour is continuous and anti-clockwise, the poles will be surrounded by clockwise circular contours such as to be excluded from the domain drawn by the blue contour. Let us note by $\Gamma_R$ the semi-circle contour, by $\Gamma_{\varepsilon,w_1}$ the circular contour of radius $\varepsilon$ surrounding the pole $w_1 = b + \varphi +$ iaγ, and by $\Gamma_{\varepsilon,w_0^+(n)}$ and $\Gamma_{\varepsilon,w_0^-(n)}$, the circular contours of radius ε surrounding the poles given by $w_0^+(n) = +w_0 + (2n + 1)i\pi$ and $w_0^-(n) = -w_0 + (2n + 1)i\pi$, respectively. One deduces that Cauchy's Integral Theorem applies and that the integral, J (Eq.B3), can be decomposed and rewritten as,

$$\int_{-R}^{+R} F(w)H(w - b - \varphi)dw + \int_{\Gamma_R} F(w)H(w - b - \varphi)dw + \int_{\Gamma_{\varepsilon,w_1}} F(w)H(w - b - \varphi)dw +$$

$$\sum_{n=0}^{n^*} \left[ \int_{\Gamma_{\varepsilon,w_0^+(n)}} F(w)H(w - b - \varphi)dw + \int_{\Gamma_{\varepsilon,w_0^-(n)}} F(w)H(w - b - \varphi)dw \right] = 0 \quad (Eq.B4)$$

Where, $n^*$ is the largest value of the set of values for n corresponding to poles being inside the blue contour. For example, $n^* = 3$ in Fig.4A. One can see clearly from Eq.B4 that the first term in the left-hand side corresponds to Eq.B1 provided one considers the limit $R \to +\infty$. In order to determine Eq.B1 we apply the method of residues namely, one considers the blue contour in the limits $R \to +\infty$ and $\varepsilon \to 0$. As $R \to +\infty$ implies $n^* \to +\infty$ one deduces as a result that Eq.B4 can be rewritten as,

$$I = -\lim_{R \to +\infty} \int_{\Gamma_R} F(w)H(w - b - \varphi)dw - \lim_{\varepsilon \to 0} \int_{\Gamma_{\varepsilon,w_1}} F(w)H(w - b - \varphi)dw -$$

$$\sum_{n=0}^{\infty} \lim_{\varepsilon \to 0} \left[ \int_{\Gamma_{\varepsilon,w_0^+(n)}} F(w)H(w - b - \varphi)dw + \int_{\Gamma_{\varepsilon,w_0^-(n)}} F(w)H(w - b - \varphi)dw \right] \quad (Eq.B5)$$

Where I is given by Eq.B2. To estimate the first term in the right-hand side one recall that as $\Gamma_R$ is a semicircle of radius R one can replace w by $Re^{i\theta}$ where $\theta \in ]0;\pi[$. Accordingly, this term can be rewritten as,

$$\lim_{R \to +\infty} \int_{\Gamma_R} F(w)H(w - \varphi)dw = \lim_{R \to +\infty} \int_0^\pi F(Re^{i\theta})H(Re^{i\theta} - b - \varphi)d(Re^{i\theta}) \quad (Eq.B6)$$

One will demonstrate now that Eq.B6 tends towards zero when $R \to +\infty$. Using the triangle inequalities one can bound the integral (Eq.B6) as follow,

$$0 \leq \lim_{R \to \infty} \left| \int_0^\pi F(Re^{i\theta})H(Re^{i\theta} - b - \varphi)Re^{i\theta}id\theta \right| \leq \lim_{R \to \infty} \int_0^\pi |F(Re^{i\theta})||H(Re^{i\theta} - b - \varphi)|Rd\theta$$

$$(Eq.B7)$$

Recalling $|F(Re^{i\theta})| = \left| \frac{sh(Re^{i\theta})}{ch(w_0)+ch(Re^{i\theta})} \right| = \frac{|sh(Re^{i\theta})|}{|ch(w_0)+ch(Re^{i\theta})|} \leq \frac{|sh(Re^{i\theta})|}{||ch(w_0)|-|ch(Re^{i\theta})||}$, one deduces then: $\lim_{R \to +\infty} |F(Re^{i\theta})| \sim 1.$ Similarly, recalling that $|H(Re^{i\theta} - b - \varphi)| =$

$\frac{a\gamma}{\pi} \frac{1}{|(a\gamma)^2+(Re^{i\theta}-b-\varphi)^2|} \leq \frac{a\gamma}{\pi} \frac{1}{|(a\gamma)^2-|(Re^{i\theta}-b-\varphi)^2||}$, one deduces: $\lim_{R\to+\infty} |H(Re^{i\theta}-b-\varphi)| \sim \frac{a\gamma}{\pi R^2}$.

As a result, in the limit $R \to +\infty$ Eq.B7 can be rewritten as,

$$0 \leq \lim_{R\to\infty} \left|\int_0^\pi F(Re^{i\theta})H(Re^{i\theta}-b-\varphi)Re^{i\theta}id\theta\right| \leq \lim_{R\to\infty} \frac{a\gamma}{R^1} \sim 0 \qquad \text{(Eq.B8)}$$

Eq.B8 demonstrates that the integral taken over the semicircle is null, i.e., $\lim_{R\to+\infty}\int_{\Gamma_R} F(w)H(w-b-\varphi)dw \sim 0$. Note that in Eq.B8 no information is visible regarding the semicircle used. That is to say that a similar result is expected also for the semicircle drawn on the red contour (Fig.4A).

The second right-hand side term from Eq.B5 can be determined by recalling that any complex number on the small blue circle of radius $\varepsilon$ surrounding the pole $w_1 = \varphi + ia\gamma$ can be written as $w_1 + \varepsilon e^{i\theta}$, with $\theta \in \;]0; -2\pi[$ as a result of the clockwise angular direction. One deduces then that this second right-hand side term can be rewritten as,

$$\lim_{\varepsilon\to 0}\int_{\Gamma_{\varepsilon,w_1}} F(w)H(w-b-\varphi)dw = \lim_{\varepsilon\to 0}\int_0^{-2\pi} F(w_1+\varepsilon e^{i\theta})H(w_1+\varepsilon e^{i\theta}-b-\varphi)\varepsilon e^{i\theta}id\theta$$

(Eq.B9)

To estimate the functions F and H in Eq.B9, one starts by noting that as long as $w_1 \neq w_0^\pm(n)$ for any value of n, then $\lim_{\varepsilon\to 0} F(w_1+\varepsilon e^{i\theta}) = \lim_{\varepsilon\to 0} \frac{sh(w_1+\varepsilon e^{i\theta})}{ch(w_0)+ch(w_1+\varepsilon e^{i\theta})} \sim F(w_1)$ while $\lim_{\varepsilon\to 0} H(w_1+\varepsilon e^{i\theta}-b-\varphi) = \lim_{\varepsilon\to 0} \frac{a\gamma}{\pi} \frac{1}{(a\gamma)^2+(ia\gamma+\varepsilon e^{i\theta})^2} \sim \frac{1}{2i\pi}\frac{1}{\varepsilon e^{i\theta}}$. Replacing these leading order relations into Eq.B9 one finds,

$$\lim_{\varepsilon\to 0}\int_0^{-2\pi} F(w_1+\varepsilon e^{i\theta})H(w_1+\varepsilon e^{i\theta}-b-\varphi)\varepsilon e^{i\theta}id\theta \sim -F(w_1) \qquad \text{(Eq.B10)}$$

Let now concentrate on the poles given by $w_0^\pm(n)$. Recalling that any complex number on the small blue circles of radius $\varepsilon$ surrounding the poles $w_0^\pm(n)$ can be written as $w_0^\pm(n) + \varepsilon e^{i\theta}$, with $\theta \in \;]0; -2\pi[$ as a result of the clockwise angular direction. One deduces,

$$\lim_{\varepsilon\to 0}\left[\int_0^{-2\pi} F(w_0^\pm(n)+\varepsilon e^{i\theta})H(w_0^\pm(n)+\varepsilon e^{i\theta}-b-\varphi)\varepsilon e^{i\theta}id\theta\right] \qquad \text{(Eq.B11)}$$

To estimate the limit $\varepsilon \to 0$ one recalls that, $F\left(w_0^{\pm}(n) + \varepsilon e^{i\theta}\right) = \frac{sh(w_0^{\pm}(n)+\varepsilon e^{i\theta})}{ch(w_0)+ch(w_0^{\pm}(n)+\varepsilon e^{i\theta})}$. As hyperbolic functions are $2i\pi$-periodic and $w_0^{\pm}(n) = \pm w_0 + (2n + 1)i\pi$, one can rewrite, F, as $F\left(w_0^{\pm}(n) + \varepsilon e^{i\theta}\right) = \frac{sh(\pm w_0 + \varepsilon e^{i\theta}+i\pi)}{ch(w_0)+ch(\pm w_0+\varepsilon e^{i\theta}+i\pi)}$. Developing the hyperbolic functions one can also write F under the form, $F\left(w_0^{\pm}(n) + \varepsilon e^{i\theta}\right) = \frac{-sh(\pm w_0+\varepsilon e^{i\theta})}{ch(w_0)-ch(\pm w_0+\varepsilon e^{i\theta})}$. Let now consider the chain rule method and develop the hyperbolic terms in the limit $\varepsilon \to 0$, one obtains $\lim_{\varepsilon \to 0} F\left(w_0^{\pm}(n) + \varepsilon e^{i\theta}\right) \sim \lim_{\varepsilon \to 0} \frac{-sh(\pm w_0)-ch(\pm w_0)\varepsilon e^{i\theta}}{ch(w_0)-ch(\pm w_0)-sh(\pm w_0)\varepsilon e^{i\theta}} \sim \frac{1}{\varepsilon e^{i\theta}}$ as a leading order. Estimating the function H in the limit $\varepsilon \to 0$ is trivial as provided $w_0^{\pm}(n) \neq w_1$ one finds, $\lim_{\varepsilon \to 0} H\left(w_0^{\pm}(n) + \varepsilon e^{i\theta} - b - \varphi\right) \sim H\left(w_0^{\pm}(n) - b - \varphi\right)$.

As a result,

$$\lim_{\varepsilon \to 0} \left[\int_0^{-2\pi} F\left(w_0^{\pm}(n) + \varepsilon e^{i\theta}\right) H\left(w_0^{\pm}(n) + \varepsilon e^{i\theta} - b - \varphi\right) \varepsilon e^{i\theta} i d\theta\right] \sim -2i\pi H\left(w_0^{\pm}(n) - b - \varphi\right)$$

(Eq.B12)

Replacing Eq.B12, Eq.B10 and Eq.B8 into Eq.B5 one finds finally,

$I \sim F(w_1) + 2i\pi \sum_{n=0}^{\infty} H(w_0^+(n) - b - \varphi) + H(w_0^-(n) - b - \varphi)$  (Eq.B13)

Eq.B13 is valid for the blue contour. However using the red contour to determine Eq.B1 is also a possibility. The only differences between the blue and the red contour concerns the poles that are conjugated elements for the red contour and the way they are surrounded by anticlockwise circles noted as opposed to clockwise circle in the blue contour. Given that the poles in the red contour are defined by, $\overline{w}_1 = b + \varphi - ia\gamma$, $\overline{w}_0^+(n) = +w_0 - (2n + 1)i\pi$ and $\overline{w}_0^-(n) = -w_0 - (2n + 1)i\pi$, the red contour leads to a determination of Eq.B1 under the form,

$I \sim F(\overline{w}_1) - 2i\pi \sum_{n=0}^{\infty} H(\overline{w}_0^+(n) - b - \varphi) + H(\overline{w}_0^-(n) - b - \varphi)$  (Eq.B14)

As the conservation of genetic microstates imposes that both Eq.B13 and Eq.B14 be valid and identical to $\omega_0 th(\varphi)$ the relations Eq.20e and Eq.20i given in the main text when $ch(w_0) \geq 1$ are therefore fully justified.

**Appendix C: Cauchy's Integral Theorem applied to genetic when:** $1 > \left(\frac{1-\omega_0}{\omega_0}\right) \text{ch}(\varphi) = \text{ch}(iw_0) > 0.$

The objective is to use Cauchy Integral Theorem in the complex map to resolve the integral given by,

$$I = \int_{-\infty}^{+\infty} \frac{\text{sh}(w)}{\text{ch}(iw_0)+\text{ch}(w)} \frac{a\gamma}{\pi} \frac{1}{(a\gamma)^2+(w-b-\varphi)^2} dw \qquad \text{(Eq.C1)}$$

To remain coherent with the notations in the main text, we shall rewrite the integral as,

$$I = \int_{-\infty}^{+\infty} F(w)H(w-b-\varphi)dw \qquad \text{(Eq.C2)}$$

As said above, Cauchy Integral Theorem used in the complex map stipulates that provided that a continuous closed contour is drawn in such a way to exclude the poles of the integrand, then the integral is null on the continuous contour chosen. Consequently, assuming now that $w$ is a complex number, and considering a continuous closed contour noted $C$ in the complex map, the integral one aims to resolve in the complex map is,

$$J = \oint_C F(w)H(w-b-\varphi)dw \qquad \text{(Eq.C3)}$$

Note that $J = 0$ if the closed contour excludes the poles of the integrand. Using the complex formulation $\left(\frac{1-\omega_0}{\omega_0}\right)\text{ch}(\varphi) = \text{ch}(iw_0)$, the poles are presented as black dots in Fig.4B and depending on the sign of the poles' imaginary part, two contours can be draw that are represented by the blue and red contours.

Let us concentrate on the blue contour from Fig.4B. This continuous blue contour is the sum of a set of elements including the large semi-circle, the diameter of the semi-circle as well as the small blue contours surrounding the poles. Note that as the overall contour is continuous and anti-clockwise, the poles will be surrounded by clockwise circular contours such as to be excluded from the domain drawn by the blue contour in the complex map. Let us note by $\Gamma_R$ the semi-circle contour, by $\Gamma_{\varepsilon,w_1}$ the circular contour of radius $\varepsilon$ surrounding the pole $w_1 = b + \varphi + ia\gamma$, and by $\Gamma_{\varepsilon,w_0(n)}$ the circular contours of radius $\varepsilon$ surrounding the poles given by $w_0(n) = +iw_0 + (2n+1)i\pi$, respectively. One deduces that Cauchy's Integral Theorem applies and that the integral, $J$ (Eq.C3), can be decomposed and rewritten as, $\int_{-R}^{+R} F(w)H(w - b - \varphi)dw + \int_{\Gamma_R} F(w)H(w - b - \varphi)dw + \int_{\Gamma_{\varepsilon,w_1}} F(w)H(w - b - \varphi)dw +$

$\sum_{n=0}^{n^*} \int_{\Gamma_{\varepsilon,w_0(n)}} F(w)H(w - b - \varphi)dw = 0$ (Eq.C4)

Where, $n^*$ is the largest value of the set of values for n corresponding to poles being inside the blue contour. For example, $n^* = 3$ in Fig.4B. One can see from Eq.C4 that first left-hand side term correspond to Eq.C1 provided one considers the limit $R \to +\infty$. In order to determine Eq.C1 we apply the method of residues namely, one considers the blue contour using in the limits $R \to +\infty$ and $\varepsilon \to 0$. As $R \to +\infty$ implies $n^* \to +\infty$ one deduces as a result that Eq.B4 can be rewritten as,

$I = - \lim_{R \to +\infty} \int_{\Gamma_R} F(w)H(w - b - \varphi)dw - \lim_{\varepsilon \to 0} \int_{\Gamma_{\varepsilon,w_1}} F(w)H(w - b - \varphi)dw -$

$\sum_{n=0}^{\infty} \lim_{\varepsilon \to 0} \int_{\Gamma_{\varepsilon,w_0(n)}} F(w)H(w - b - \varphi)dw$ (Eq.C5)

The first term was estimated in Eq.B8 and was found to be null. The second right-hand side term in Eq.C5 can be determined by recalling that any complex number on the small blue circle of radius $\varepsilon$ surrounding the pole $w_1 = b + \varphi + ia\gamma$ can be written as $w_1 + \varepsilon e^{i\theta}$, with $\theta \in ]0; -2\pi[$ as a result of the clockwise angular direction. One deduces then that this second right-hand side term can be rewritten as,

$\lim_{\varepsilon \to 0} \int_{\Gamma_{\varepsilon,w_1}} F(w)H(w - b - \varphi)dw = \lim_{\varepsilon \to 0} \int_0^{-2\pi} F(w_1 + \varepsilon e^{i\theta})H(w_1 + \varepsilon e^{i\theta} - b - \varphi)\varepsilon e^{i\theta} id\theta$

(Eq.C6)

To estimate the functions F and H in Eq.C6, one starts by noting that as long as $w_1 \neq w_0(n)$ for any value of n, then $\lim_{\varepsilon \to 0} F(w_1 + \varepsilon e^{i\theta}) = \lim_{\varepsilon \to 0} \frac{sh(w_1+\varepsilon e^{i\theta})}{ch(iw_0)+ch(w_1+\varepsilon e^{i\theta})} \sim F(w_1)$ while $\lim_{\varepsilon \to 0} H(w_1 + \varepsilon e^{i\theta} - b - \varphi) = \lim_{\varepsilon \to 0} \frac{a\gamma}{\pi} \frac{1}{(a\gamma)^2+(ia\gamma+\varepsilon e^{i\theta})^2} \sim \frac{1}{2i\pi} \frac{1}{\varepsilon e^{i\theta}}$. Replacing these leading order relations into Eq.C6 one finds,

$\lim_{\varepsilon \to 0} \int_0^{-2\pi} F(w_1 + \varepsilon e^{i\theta})H(w_1 + \varepsilon e^{i\theta} - b - \varphi)\varepsilon e^{i\theta} id\theta \sim - F(w_1)$ (Eq.C7)

Let now concentrate on the poles given by $w_0(n)$. Recalling that any complex number on the contour $\Gamma_{\varepsilon,w_0(n)}$, i.e., small blue circles of radius $\varepsilon$ surrounding the poles $w_0(n)$, can be written as $w_0(n) + \varepsilon e^{i\theta}$, with $\theta \in ]0; -2\pi[$ as a result of the clockwise angular direction. One deduces,

$$\lim_{\varepsilon \to 0} \left[ \int_0^{-2\pi} F(w_0(n) + \varepsilon e^{i\theta}) H(w_0(n) + \varepsilon e^{i\theta} - b - \varphi) \varepsilon e^{i\theta} i d\theta \right] \quad \text{(Eq.C8)}$$

To estimate the limit $\varepsilon \to 0$ one recalls that, $F(w_0(n) + \varepsilon e^{i\theta}) = \frac{sh(w_0(n)+\varepsilon e^{i\theta})}{ch(iw_0)+ch(w_0(n)+\varepsilon e^{i\theta})}$. As hyperbolic functions are $2i\pi$-periodic and $w_0(n) = +iw_0 + (2n+1)i\pi$, one can rewrite, F, as $F(w_0(n) + \varepsilon e^{i\theta}) = \frac{sh(iw_0+\varepsilon e^{i\theta}+i\pi)}{ch(iw_0)+ch(iw_0+\varepsilon e^{i\theta}+i\pi)}$. Developing the hyperbolic functions one can also write F under the form, $F(w_0(n) + \varepsilon e^{i\theta}) = \frac{-sh(iw_0+\varepsilon e^{i\theta})}{ch(iw_0)-ch(iw_0+\varepsilon e^{i\theta})}$. Let now consider the chain rule method and develop the hyperbolic terms in the limit $\varepsilon \to 0$, one obtains $\lim_{\varepsilon \to 0} F(w_0(n) + \varepsilon e^{i\theta}) \sim \lim_{\varepsilon \to 0} \frac{-sh(iw_0)-ch(iw_0)\varepsilon e^{i\theta}}{ch(iw_0)-ch(iw_0)-sh(iw_0)\varepsilon e^{i\theta}} \sim \frac{1}{\varepsilon e^{i\theta}}$ as a leading order. Estimating the function H in the limit $\varepsilon \to 0$ is trivial as provided $w_0(n) \neq w_1$ one finds, $\lim_{\varepsilon \to 0} H(w_0(n) + \varepsilon e^{i\theta} - b - \varphi) \sim H(w_0(n) - b - \varphi)$.

As a result,

$$\lim_{\varepsilon \to 0} \left[ \int_0^{-2\pi} F(w_0(n) + \varepsilon e^{i\theta}) H(w_0(n) + \varepsilon e^{i\theta} - b - \varphi) \varepsilon e^{i\theta} i d\theta \right] \sim -2i\pi H(w_0(n) - b - \varphi)$$

(Eq.C9)

Replacing Eq.B8, Eq.C7 and Eq.C9 into Eq.C5 one finds finally,

$$I \sim F(w_1) + 2i\pi \sum_{n=0}^{\infty} H(w_0(n) - b - \varphi) \quad \text{(Eq.C10)}$$

Eq.C10 is valid for the blue contour. However using the red contour to determine Eq.C1 is also a possibility. The only differences between the blue and the red contour concerns the poles that are now conjugated elements for the red contour and the way they are surrounded by anticlockwise circles noted $\Gamma_{\varepsilon,w_0(n)}$ as opposed to clockwise circle in the blue contour. Given that the poles in the red contour are defined by, $\bar{w}_1 = b + \varphi - ia\gamma$, $\bar{w}_0(n) = -iw_0 - (2n+1)i\pi$, the red contour leads to a determination of Eq.C1 under the form,

$$I \sim F(\bar{w}_1) - 2i\pi \sum_{n=0}^{\infty} H(\bar{w}_0(n) - b - \varphi) \quad \text{(Eq.C11)}$$

As the conservation of genetic microstates imposes that both Eq.C10 and Eq.C11 be valid and identical to $\omega_0 th(\varphi)$.

**Appendix D: Cauchy's Integral Theorem applied to genetic when:** $\left(\frac{1-\omega_0}{\omega_0}\right) ch(\varphi) \geq 1$.

The objective is to use Cauchy Integral Theorem in the complex map to resolve the integral given by,

$$I = \int_{-\infty}^{+\infty} \frac{sh(w)}{ch(w_0)+ch(w)} \frac{a\gamma}{\pi} \frac{1}{(a\gamma)^2+(w-b-\varphi)^2} dw \qquad (Eq.D1)$$

To remain coherent with the notations in the main text, we shall rewrite the integral as,

$$I = \int_{-\infty}^{+\infty} \tilde{F}(w) H(w-b-\varphi) dw \qquad (Eq.D2)$$

As said above, Cauchy Integral Theorem used in the complex map stipulates that provided that a continuous closed contour is drawn in such a way to exclude the poles of the integrand, then the integral is null on the continuous contour chosen. Consequently, assuming now that $w$ is a complex number, and considering a continuous closed contour noted $C$ in the complex map, the integral one aims to resolve in the complex map is,

$$J = \oint_C \tilde{F}(w) H(w-b-\varphi) dw \qquad (Eq.D3)$$

Note that $J = 0$ if the closed contour excludes the poles of the integrand. Using the complex formulation $\left(\frac{1-\omega_0}{\omega_0}\right) ch(\varphi) = ch(w_0)$, the poles are presented as black dots in Fig.4A and depending on the sign of the poles' imaginary part, two contours can be draw that are represented by the blue and red contours.

Let us concentrate on the blue contour from Fig.4A. This continuous blue contour is the sum of a set of elements including the large semi-circle, the diameter of the semi-circle as well as the small blue contours surrounding the poles. Note that as the overall contour is continuous and anti-clockwise, the poles will be surrounded by clockwise circular contours such as to be excluded from the domain drawn by the blue contour in the complex map. Let us note by $\Gamma_R$ the semi-circle contour, by $\Gamma_{\varepsilon,w_1}$ the circular contour of radius $\varepsilon$ surrounding the pole $w_1 = b + \varphi + ia\gamma$, and by $\Gamma_{\varepsilon,w_0^+(n)}$ and $\Gamma_{\varepsilon,w_0^-(n)}$ the circular contours of radius $\varepsilon$ surrounding the poles given by $w_0^+(n) = +w_0 + (2n+1)i\pi$ and $w_0^-(n) = -w_0 + (2n+1)i\pi$, respectively. One deduces that Cauchy's Integral Theorem applies and that the integral, $J$ (Eq.D3), can be decomposed and rewritten as,

$$\int_{-R}^{+R} \tilde{F}(w)H(w-b-\varphi)dw + \int_{\Gamma_R} \tilde{F}(w)H(w-b-\varphi)dw + \int_{\Gamma_{\varepsilon,w_1}} \tilde{F}(w)H(w-b-\varphi)dw +$$

$$\sum_{n=0}^{n^*} \left[ \int_{\Gamma_{\varepsilon,w_0^+(n)}} \tilde{F}(w)H(w-b-\varphi)dw + \int_{\Gamma_{\varepsilon,w_0^-(n)}} \tilde{F}(w)H(w-b-\varphi)dw \right] = 0 \quad \text{(Eq.D4)}$$

Where, $n^*$ is the largest value of the set of values for n corresponding to poles being inside the blue contour. For example, $n^* = 3$ in Fig.4A. One can see from Eq.D4 that first left-hand side term correspond to Eq.D1 provided one considers the limit $R \to +\infty$. In order to determine Eq.D1 we apply the method of residues namely, one considers the blue contour using in the limits $R \to +\infty$ and $\varepsilon \to 0$. As $R \to +\infty$ implies $n^* \to +\infty$ one deduces as a result that Eq.B4 can be rewritten as,

$$I = -\lim_{R \to +\infty} \int_{\Gamma_R} \tilde{F}(w)H(w-b-\varphi)dw - \lim_{\varepsilon \to 0} \int_{\Gamma_{\varepsilon,w_1}} \tilde{F}(w)H(w-b-\varphi)dw -$$

$$\sum_{n=0}^{\infty} \lim_{\varepsilon \to 0} \left[ \int_{\Gamma_{\varepsilon,w_0^+(n)}} \tilde{F}(w)H(w-b-\varphi)dw + \int_{\Gamma_{\varepsilon,w_0^-(n)}} \tilde{F}(w)H(w-b-\varphi)dw \right] \quad \text{(Eq.D5)}$$

The first term was estimated in Eq.B8 and was found to be null. The second right-hand side term in Eq.D5 can be determined by recalling that any complex number on the small blue circle of radius $\varepsilon$ surrounding the pole $w_1 = b + \varphi + ia\gamma$ can be written as $w_1 + \varepsilon e^{i\theta}$, with $\theta \in\ ]0; -2\pi[$ as a result of the clockwise angular direction. One deduces then that this second right-hand side term can be rewritten as,

$$\lim_{\varepsilon \to 0} \int_{\Gamma_{\varepsilon,w_1}} \tilde{F}(w)H(w-b-\varphi)dw = \lim_{\varepsilon \to 0} \int_0^{-2\pi} \tilde{F}(w_1 + \varepsilon e^{i\theta})H(w_1 + \varepsilon e^{i\theta} - b - \varphi)\varepsilon e^{i\theta} id\theta$$

(Eq.D6)

To estimate the functions $\tilde{F}$ and H in Eq.D6, one starts by noting that as long as $w_1 \neq \pm w_0 + (2n+1)i\pi$ for any value of n, then $\lim_{\varepsilon \to 0} \tilde{F}(w_1 + \varepsilon e^{i\theta}) = \lim_{\varepsilon \to 0} \frac{ch(w_1+\varepsilon e^{i\theta})}{ch(w_0)+ch(w_1+\varepsilon e^{i\theta})} \sim \tilde{F}(w_1)$ while $\lim_{\varepsilon \to 0} H(w_1 + \varepsilon e^{i\theta} - b - \varphi) = \lim_{\varepsilon \to 0} \frac{a\gamma}{\pi} \frac{1}{(a\gamma)^2+(ia\gamma+\varepsilon e^{i\theta})^2} \sim \frac{1}{2i\pi} \frac{1}{\varepsilon e^{i\theta}}$. Replacing these leading order relations into Eq.D6 one finds,

$$\lim_{\varepsilon \to 0} \int_0^{-2\pi} \tilde{F}(w_1 + \varepsilon e^{i\theta})H(w_1 + \varepsilon e^{i\theta} - b - \varphi)\varepsilon e^{i\theta} id\theta \sim -F(w_1) \quad \text{(Eq.D7)}$$

Let now concentrate on the poles given by $w_0^\pm(n)$. Recalling that any complex number on the small blue circles of radius $\varepsilon$ surrounding the poles $w_0^\pm(n)$ can be written as $w_0^\pm(n) + \varepsilon e^{i\theta}$, with $\theta \in\ ]0; -2\pi[$ as a result of the clockwise angular direction. One deduces,

$$\lim_{\varepsilon \to 0} \left[\int_0^{-2\pi} \tilde{F}(w_0^\pm(n) + \varepsilon e^{i\theta}) H(w_0^\pm(n) + \varepsilon e^{i\theta} - b - \varphi) \varepsilon e^{i\theta} i d\theta\right] \qquad (Eq.D8)$$

To estimate the limit $\varepsilon \to 0$ one recalls that, $F(w_0^\pm(n) + \varepsilon e^{i\theta}) = \frac{ch(w_0^\pm(n) + \varepsilon e^{i\theta})}{ch(w_0) + ch(w_0^\pm(n) + \varepsilon e^{i\theta})}$. As hyperbolic functions are $2i\pi$-periodic and $w_0^\pm(n) = \pm w_0 + (2n+1)i\pi$, one can rewrite, $\tilde{F}$, as $\tilde{F}(w_0^\pm(n) + \varepsilon e^{i\theta}) = \frac{ch(\pm w_0 + \varepsilon e^{i\theta} + i\pi)}{ch(w_0) + ch(\pm w_0 + \varepsilon e^{i\theta} + i\pi)}$. Developing the hyperbolic functions one can also write F under the form, $\tilde{F}(w_0^\pm(n) + \varepsilon e^{i\theta}) = \frac{-ch(\pm w_0 + \varepsilon e^{i\theta})}{ch(w_0) - ch(\pm w_0 + \varepsilon e^{i\theta})}$. Let now consider the chain rule method and develop the hyperbolic terms in the limit $\varepsilon \to 0$, one obtains $\lim_{\varepsilon \to 0} \tilde{F}(w_0^\pm(n) + \varepsilon e^{i\theta}) \sim \lim_{\varepsilon \to 0} \frac{-ch(\pm w_0)}{ch(w_0) - ch(\pm w_0) - sh(\pm w_0)\varepsilon e^{i\theta}} \sim \frac{1}{th(\pm w_0)} \frac{1}{\varepsilon e^{i\theta}}$ as a leading order. Estimating the function H in the limit $\varepsilon \to 0$ is trivial as provided $w_0^\pm(n) \neq w_1$ one finds, $\lim_{\varepsilon \to 0} H(w_0^\pm(n) + \varepsilon e^{i\theta} - b - \varphi) \sim H(w_0^\pm(n) - b - \varphi)$.

As a result,

$$\lim_{\varepsilon \to 0}\left[\int_0^{-2\pi} \tilde{F}(w_0^\pm(n) + \varepsilon e^{i\theta}) H(w_0^\pm(n) + \varepsilon e^{i\theta} - b - \varphi) \varepsilon e^{i\theta} i d\theta\right] \sim -2i\pi \frac{1}{th(\pm w_0)} H(w_0^\pm(n) - b - \varphi)$$

(Eq.D9)

Replacing Eq.D9, Eq.D7 and Eq.B8 into Eq.D5 one finds finally,

$$I \sim \tilde{F}(w_1) + 2i\pi \frac{1}{th(+w_0)} \sum_{n=0}^{\infty} [H(w_0^+(n) - b - \varphi) - H(w_0^-(n) - b - \varphi)] \qquad (Eq.D10)$$

Eq.D10 is valid for the blue contour. However, using the red contour to determine Eq.D1 is also a possibility. The only differences between the blue and the red contour concerns the poles that are now conjugated elements for the red contour and the way they are surrounded by anticlockwise circles as opposed to clockwise circle in the blue contour. Given that the poles in the red contour are defined by, $\bar{w}_1 = b + \varphi - ia\gamma$, $\bar{w}_0^+(n) = +w_0 - (2n+1)i\pi$ and

$\overline{w}_0^-(n) = -w_0 - (2n+1)i\pi$, the red contour leads to a determination of Eq.B1 under the form,

$$I \sim \tilde{F}(\overline{w}_1) - 2i\pi \frac{1}{\text{th}(+w_0)} \sum_{n=0}^{\infty} [H(\overline{w}_0^+(n) - b - \varphi) - H(\overline{w}_0^-(n) - b - \varphi)] \qquad (Eq.D11)$$

As the conservation of genetic microstates imposes that Eq.D10 and Eq.11 are identical to $\omega_0$.

**Appendix E: Cauchy's Integral Theorem applied to genetic when: $1 > \left(\frac{1-\omega_0}{\omega_0}\right) \text{ch}(\varphi) = \text{ch}(iw_0) > 0$.**

The objective is to use Cauchy Integral Theorem in the complex map to resolve the integral given by,

$$I = \int_{-\infty}^{+\infty} \frac{\text{ch}(w)}{\text{ch}(iw_0) + \text{ch}(w)} \frac{a\gamma}{\pi} \frac{1}{(a\gamma)^2 + (w-b-\varphi)^2} dw \qquad (Eq.E1)$$

To remain coherent with the notations in the main text, we shall rewrite the integral as,

$$I = \int_{-\infty}^{+\infty} \tilde{F}(w) H(w - b - \varphi) dw \qquad (Eq.E2)$$

As said above, Cauchy Integral Theorem used in the complex map stipulates that provided that a continuous closed contour is drawn in such a way to exclude the poles of the integrand, then the integral is null on the continuous contour chosen. Consequently, assuming now that $w$ is a complex number, and considering a continuous closed contour noted $C$ in the complex map, the integral one aims to resolve in the complex map is,

$$J = \oint_C \tilde{F}(w) H(w - b - \varphi) dw \qquad (Eq.E3)$$

Note that $J = 0$ if the closed contour excludes the poles of the integrand. Using the complex formulation $\left(\frac{1-\omega_0}{\omega_0}\right) \text{ch}(\varphi) = \text{ch}(iw_0)$, the poles are presented as black dots in Fig.4B and depending on the sign of the poles' imaginary part, two contours can be draw that are represented by the blue and red contours.

Let us concentrate on the blue contour from Fig.4B. This continuous blue contour is the sum of a set of elements including the large semi-circle, the diameter of the semi-circle as well as the small blue contours surrounding the poles. Note that as the overall contour is continuous and anti-clockwise, the poles will be surrounded by clockwise circular contours such as to be excluded from the domain drawn by the blue contour in the complex map. Let us note by $\Gamma_R$ the semi-circle contour, by $\Gamma_{\varepsilon,w_1}$ the circular contour of radius $\varepsilon$ surrounding the pole $w_1 = b + \varphi + ia\gamma$, and by $\Gamma_{\varepsilon,w_0(n)}$ the circular contours of radius $\varepsilon$ surrounding the poles given by $w_0(n) = +iw_0 + (2n+1)i\pi$, respectively. One deduces that Cauchy's Integral Theorem applies and that the integral, J (Eq.E3), can be decomposed and rewritten as,

$$\int_{-R}^{+R} \tilde{F}(w)H(w-b-\varphi)dw + \int_{\Gamma_R} \tilde{F}(w)H(w-b-\varphi)dw + \int_{\Gamma_{\varepsilon,w_1}} \tilde{F}(w)H(w-b-\varphi)dw +$$

$$\sum_{n=0}^{n^*} \int_{\Gamma_{\varepsilon,w_0(n)}} \tilde{F}(w)H(w-b-\varphi)dw = 0 \qquad (Eq.E4)$$

Where, $n^*$ is the largest value of the set of values for n corresponding to poles being inside the blue contour. For example, $n^* = 3$ in Fig.4B. One can see from Eq.E4 that first left-hand side term correspond to Eq.E1 provided one considers the limit $R \to +\infty$. In order to determine Eq.E1 we apply the method of residues namely, one considers the blue contour using in the limits $R \to +\infty$ and $\varepsilon \to 0$. As $R \to +\infty$ implies $n^* \to +\infty$ one deduces as a result that Eq.E4 can be rewritten as,

$$I = -\lim_{R\to+\infty} \int_{\Gamma_R} \tilde{F}(w)H(w-b-\varphi)dw - \lim_{\varepsilon\to 0} \int_{\Gamma_{\varepsilon,w_1}} \tilde{F}(w)H(w-b-\varphi)dw -$$

$$\sum_{n=0}^{\infty} \lim_{\varepsilon\to 0} \int_{\Gamma_{\varepsilon,w_0(n)}} \tilde{F}(w)H(w-b-\varphi)dw \qquad (Eq.E5)$$

The first term was estimated in Eq.B8 and was found to be null. The second right-hand side term in Eq.E5 can be determined by recalling that any complex number on the small blue circle of radius $\varepsilon$ surrounding the pole $w_1 = b + \varphi + ia\gamma$ can be written as $w_1 + \varepsilon e^{i\theta}$, with $\theta \in ]0; -2\pi[$ as a result of the clockwise angular direction. One deduces then that this second right-hand side term can be rewritten as,

$$\lim_{\varepsilon\to 0} \int_{\Gamma_{\varepsilon,w_1}} \tilde{F}(w)H(w-b-\varphi)dw = \lim_{\varepsilon\to 0} \int_0^{-2\pi} \tilde{F}(w_1 + \varepsilon e^{i\theta})H(w_1 + \varepsilon e^{i\theta} - b - \varphi)\varepsilon e^{i\theta}id\theta$$

(Eq.E6)

To estimate the functions $\tilde{F}$ and H in Eq.C6, one starts by noting that as long as $w_1 \neq w_0(n)$ for any value of n, then $\lim_{\varepsilon\to 0} \tilde{F}(w_1 + \varepsilon e^{i\theta}) = \lim_{\varepsilon\to 0} \frac{ch(w_1+\varepsilon e^{i\theta})}{ch(iw_0)+ch(w_1+\varepsilon e^{i\theta})} \sim \tilde{F}(w_1)$ while $\lim_{\varepsilon \to 0} H(w_1 + \varepsilon e^{i\theta} - b - \varphi) = \lim_{\varepsilon \to 0} \frac{a\gamma}{\pi} \frac{1}{(a\gamma)^2 + (ia\gamma + \varepsilon e^{i\theta})^2} \sim \frac{1}{2i\pi} \frac{1}{\varepsilon e^{i\theta}}$. Replacing these leading order relations into Eq.E6 one finds,

$$\lim_{\varepsilon \to 0} \int_0^{-2\pi} \tilde{F}(w_1 + \varepsilon e^{i\theta}) H(w_1 + \varepsilon e^{i\theta} - b - \varphi) \varepsilon e^{i\theta} i d\theta \sim -F(w_1) \qquad (Eq.E7)$$

Let now concentrate on the poles given by $w_0(n)$. Recalling that any complex number on the contour $\Gamma_{\varepsilon, w_0(n)}$, i.e., small blue circles of radius $\varepsilon$ surrounding the poles $w_0(n)$, can be written as $w_0(n) + \varepsilon e^{i\theta}$, with $\theta \in ]0; -2\pi[$ as a result of the clockwise angular direction. One deduces,

$$\lim_{\varepsilon \to 0} \left[ \int_0^{-2\pi} \tilde{F}(w_0(n) + \varepsilon e^{i\theta}) H(w_0(n) + \varepsilon e^{i\theta} - b - \varphi) \varepsilon e^{i\theta} i d\theta \right] \qquad (Eq.E8)$$

To estimate the limit $\varepsilon \to 0$ one recalls that, $\tilde{F}(w_0(n) + \varepsilon e^{i\theta}) = \frac{ch(w_0(n) + \varepsilon e^{i\theta})}{ch(iw_0) + ch(w_0(n) + \varepsilon e^{i\theta})}$. As hyperbolic functions are $2i\pi$-periodic and $w_0(n) = +iw_0 + (2n+1)i\pi$, one can rewrite, $\tilde{F}$, as $\tilde{F}(w_0(n) + \varepsilon e^{i\theta}) = \frac{ch(iw_0 + \varepsilon e^{i\theta} + i\pi)}{ch(iw_0) + ch(iw_0 + \varepsilon e^{i\theta} + i\pi)}$. Developing the hyperbolic functions one can also write $\tilde{F}$ under the form, $\tilde{F}(w_0(n) + \varepsilon e^{i\theta}) = \frac{-ch(iw_0 + \varepsilon e^{i\theta})}{ch(iw_0) - ch(iw_0 + \varepsilon e^{i\theta})}$. Let now consider the chain rule method and develop the hyperbolic terms in the limit $\varepsilon \to 0$, one obtains then $\lim_{\varepsilon \to 0} \tilde{F}(w_0(n) + \varepsilon e^{i\theta}) \sim \lim_{\varepsilon \to 0} \frac{-ch(iw_0) - sh(iw_0) \varepsilon e^{i\theta}}{ch(iw_0) - ch(iw_0) - sh(iw_0) \varepsilon e^{i\theta}} \sim \frac{1}{th(iw_0)} \frac{1}{\varepsilon e^{i\theta}}$ as a leading order. Estimating the function H in the limit $\varepsilon \to 0$ is trivial as provided $w_0(n) \neq w_1$ one finds, $\lim_{\varepsilon \to 0} H(w_0(n) + \varepsilon e^{i\theta} - b - \varphi) \sim H(w_0(n) - b - \varphi)$.

As a result,

$$\lim_{\varepsilon \to 0} \left[ \int_0^{-2\pi} F(w_0(n) + \varepsilon e^{i\theta}) H(w_0(n) + \varepsilon e^{i\theta} - b - \varphi) \varepsilon e^{i\theta} i d\theta \right] \sim -\frac{2i\pi}{th(iw_0)} H(w_0(n) - b - \varphi)$$

(Eq.E9)

Replacing Eq.B8, Eq.E7 and Eq.E9 into Eq.E5 one finds finally,

$$I \sim F(w_1) + \frac{2i\pi}{th(iw_0)} \sum_{n=0}^{\infty} H(w_0(n) - b - \varphi) \qquad (Eq.E10)$$

Eq.E10 is valid for the blue contour. However using the red contour to determine Eq.E1 is also a possibility. The only differences between the blue and the red contour concerns the poles that are now conjugated elements for the red contour and the way they are surrounded by anticlockwise circles noted $\Gamma_{\varepsilon,\overline{w}_0(n)}$ as opposed to clockwise circle in the blue contour. Given that the poles in the red contour are defined by, $\overline{w}_1 = b + \varphi - ia\gamma$, $\overline{w}_0(n) = -iw_0 - (2n+1)i\pi$, the red contour leads to a determination of Eq.E1 under the form,

$$I \sim F(\overline{w}_1) - \frac{2i\pi}{th(iw_0)} \sum_{n=0}^{\infty} H(\overline{w}_0(n) - b - \varphi) \qquad (Eq.E11)$$

The conservation of genetic microstates imposes that both Eq.E10 and Eq.E11 be valid and identical to $\omega_0$.